\documentclass[twocolumn,floatfix]{aastex63}
\usepackage{float}
\usepackage{multirow}
\usepackage{lineno}
\usepackage{graphicx} 
\usepackage{subfigure} 
\usepackage{enumitem}
\usepackage{mathtools}
\usepackage{rotating}

\makeatletter

\newcommand{\Rmnum}[1]{\expandafter\@slowromancap\romannumeral #1@}
\makeatother

\shortauthors{R. Wang et al.}

\begin{document}

\newcommand\teff{T$_{\mathrm{eff}}$}
\newcommand\logg{log\,{\it g}}
\newcommand\kms{km\,s$^{-1}$}
\newcommand\afe{[$\alpha$/M]}

\newcommand{\secref}[1]{Section~\ref{sec:#1}}
\newcommand{\Eq}[1]{Equation~\ref{eq:#1}}
\newcommand{\Eqs}[2]{Equations~\ref{eq:#1} and~\ref{eq:#2}}
\newcommand{\Fig}[1]{Figure~\ref{#1}}
\newcommand{\csn}[0]{\textsc{cycle-starnet }}
\newcommand{\csnn}[0]{\textsc{cycle-starnet}}
\title{Stellar Parameters and Chemical Abundances Estimated from LAMOST-\Rmnum{2} DR8 MRS based on Cycle-StarNet}

\author[0000-0001-6767-2395]{Rui Wang}
\affiliation{Key Laboratory of Optical Astronomy, National Astronomical Observatories, Chinese Academy of Sciences, Beijing 100012, China}

\author[0000-0001-7865-2648]{A-Li Luo}
\affiliation{Key Laboratory of Optical Astronomy, National Astronomical Observatories, Chinese Academy of Sciences, Beijing 100012, China}
\affiliation{University of Chinese Academy of Sciences, Beijing 100049, China}

\author[0000-0003-1454-1636]{Shuo Zhang}
\affiliation{Department of Astronomy, Peking University, Beijing 100871, China}
\affiliation{Kavli institute of Astronomy and Astrophysics, Peking University, Beijing 100871, China}

\author[0000-0001-5082-9536]{Yuan-Sen Ting}
\affiliation{Research School of Astronomy $\&$ Astrophysics, Australian National University, Cotter Rd, Weston, ACT 2611, Australia}
\affiliation{School of Computing, Australian National University, Acton, ACT 2601, Australia}

\author[0000-0003-3719-1990]{Teaghan O'Briain}
\affiliation{Department of Physics and Astronomy, University of Victoria, Victoria, BC V9E 2E7, Canada}

\author{LAMOST MRS Collaboration}
\affil{Department of Astronomy, Beijing Normal University, Beijing 100875, China}
\affil{Key Laboratory of Optical Astronomy, National Astronomical Observatories, Chinese Academy of Sciences, Beijing 100012, China}
\affil{Yunnan Observatories, Chinese Academy of Sciences, Kunming 650216, China}
\affil{Kavli Institute for Astronomy and Astrophysics, Peking University, Beijing 100871, China}
\affil{School of Astronomy and Space Science, Nanjing University, Nanjing 210093, China}
\affil{Department of Physics, Hebei Normal University, Shijiazhuang 050024, China}
\affil{Department of Astronomy and Institute of Theoretical Physics and Astrophysics, Xiamen University, Xiamen 361005, China}

\correspondingauthor{A-Li Luo}
\email{lal@bao.ac.cn \& wangrui@nao.cas.cn}

\begin{abstract}

Deriving stellar atmospheric parameters and chemical abundances from stellar spectra is crucial for understanding the evolution of the Milky Way. By performing a fitting with MARCS model atmospheric theoretical synthetic spectra combined with a domain-adaptation method, we estimate the fundamental stellar parameters (\teff, \logg, [Fe/H], $v_{mic}$ and $v_{mac}$) and 11 chemical abundances for 1.38 million FGKM-type stars of the Medium-Resolution Spectroscopic Survey (MRS) from LAMOST-\Rmnum{2} DR8. The domain-adaptation method, \csnn, is employed to reduce the gap between observed and synthetic spectra, and the L-BFGS algorithm is used to search the best-fit synthetic spectra. By combining the 2MASS photometric survey data, Gaia EDR3 parallax, and MIST isochrones, the surface gravities of the stars are constrained after estimating their bolometric luminosities. The accuracy of \teff, \logg, and $[\text{Fe}/\text{H}]$ can reach 150 K, 0.11 dex and 0.15 dex, evaluated by the PASTEL catalog, asteroseismic samples, and other spectroscopic surveys. The precision of these parameters and elemental abundances ([C/Fe], [Na/Fe], [Mg/Fe], [Si/Fe], [Ca/Fe], [Ti/Fe], [Cr/Fe], [Mn/Fe], [Co/Fe], [Ni/Fe], and [Cu/Fe]) is assessed by repeated observations and validated by cluster members. For spectra with signal-to-noise (S/N) ratios greater than 10, the precision of the three stellar parameters and elemental abundances can achieve 76 K, 0.014 dex, 0.096 dex, and 0.04-0.15 dex. For spectra with S/N ratios higher than 100, the precision stabilizes at 22 K, 0.006 dex, 0.043 dex, and 0.01-0.06 dex. The full LAMOST MRS stellar properties catalog is available on-line.

\end{abstract}

\keywords{stars: atmospheres -- methods: data analysis -- techniques: spectroscopic}

\section{Introduction}

The atmospheric parameters and chemical abundances of the stars play a pivotal role in elucidating their evolutionary stages, and offer valuable information on revealing the formation history of the Milky Way. Accuracy and precision of these parameters are important statistical criteria in measurement work, and improving both of them is always the goal pursued~\citep{2019ARA&A..57..571J}. So far, it is generally considered that high-resolution (R $\geq$ 40,000) spectroscopy project, such as Gaia-ESO~\citep{2012Msngr.147...25G}, is suitable for the observation and measurement tasks of the standard benchmark stars because their fine spectral characteristics can well constrain theoretical models to provide accurate star parameters and chemical abundances. However, limited by the observation condition and exposure time, obtaining a huge collection of high-quality high-resolution spectra is very expensive. In recent years,~\citet{2016ApJ...826L..25R},~\citet{2017ApJ...843...32T},~\citet{2019ApJS..245...34X} have shown that stellar parameters and 10-20 chemical elements (with precisions of 0.01 dex to 0.1 dex) can be extracted from low-resolution (R $ \leq $ 6000) high-quality (S/N $ > $ 100) spectra based on data-driven methods, which provide theoretical and technical support for analysis of low-/medium-resolution (R from 1800 to 20000) spectra of multiple objects spectroscopic survey such as LAMOST~\citep{2015RAA....15.1095L,2020arXiv200507210L}, Subaru PFS~\citep{2014PASJ...66R...1T}, 4MOST~\citep{2019Msngr.175....3D}. These surveys can greatly expand the distribution range of the atmospheric parameters and chemical abundances, and reveal the formation history of the Milky Way by combining astrometric information~\citep{2022nature-xiang}.

The spectral fitting techniques are mostly used to derive stellar parameters by synthesizing the most similar theoretical spectra to transfer theoretical stellar labels to the observed spectra. The industrial measurement process often relies on meticulous spectral reduction, accurate theoretical spectral grids, and efficient fitting algorithms. Besides, the most important materials are the construction of the model spectra. The choice of theoretical atmosphere model (1D/LTE or 3D/N-LTE), the atomic and molecular line lists, the solar abundances composition, isotope ratio, and the parameter sampling space should be made to produce synthetic spectra according to the source selection strategies and the characteristics of the observations. The fitting techniques need to be specially designed based on spectral characteristics. The theoretical spectra and the observed spectra often show separate distributions in the high-dimensional flux space, due to the imperfection of the theoretical spectra and the observed errors caused by the observation conditions and various instrument effects.~\citet{2021ApJ...906..130O, 2020arXiv200703112O} proposed a method, \csnn, based on domain adaptation to reduce the gap between the synthetic and the observed spectra, which combines the data consistency of data-driven methods and the physical interpretability of model-driven methods.

The data-driven methods promoted cooperation between the international spectroscopic surveys, which improved the development of the measurement of the stellar parameters for low- and medium-resolution spectra.~\citet{2019ApJS..245...34X} estimated the stellar parameters and chemical abundances of 1.6 million LAMOST low-resolution spectra by transferring APOGEE and GALAH stellar labels.~\citet{2020ApJ...898...58W} analyzed millions of the low-resolution spectra to obtain several chemical elements of 5 channels using the Cannon~\citep{2015ApJ...808...16N} and revealed the distribution of the Milky Way from the perspective of chemical kinematics.~\citet{2021ApJS..253...22X,2021arXiv210802878X} also predicted the atmospheric parameters and elemental abundances of $\sim$ 100,000 hot stars largely expanding the temperature range. For cool stars, ~\citet{2021ApJS..253...45L} derived the atmospheric parameters of LAMOST M stars based on the methods SLAM~\citep{2020ApJS..246....9Z}.~\citet{2020ApJ...891...23W} and~\citet{2020A&A...644A.168G} demonstrated that data-driven methods are capable of estimating the stellar parameters and chemical abundances of the medium-resolution spectra (R $\sim$ 7500) covering a narrow band such as LAMOST DR7 and RAVE DR6. The data-driven methods learn the gradient variation between the input fluxes of spectra and the output stellar labels (or input stellar labels with output fluxes), but sometimes they learn the correlation that is superficial to the data, not physical. Besides, the extrapolation results beyond the range of training set parameters are unreliable~\citep{2021MNRAS.506..150B}.

LAMOST-\Rmnum{2} medium-resolution spectroscopic survey (LAMOST-\Rmnum{2} MRS) is the advanced project of the LAMOST-\Rmnum{1} low-resolution spectroscopic survey~\citep{2015RAA....15.1095L}. LAMOST MRS survey is expected to advance stellar physical science research, such as detailed observations of various variable stars~\citep{2020ApJS..251...15Z,2021RAA....21..292W}, planet host stars~\citep{2021ApJ...909..115C, 2021AJ....162..100C}, emission nebulae~\citep{2021RAA....21...96W, 2021RAA....21..280Z,2021RAA....21...51R}, open star clusters, young pre-main-sequence stars, etc (see ~\citet{2020arXiv200507210L} for more details). 
From the sixth Data Release, LAMOST began to release the medium-resolution (R $\sim$ 7500) spectra by the LAMOST data-process pipeline, two catalogs of the stellar parameters: one is by the LAMOST stellar parameters pipeline~\citep[LASP;][]{2015RAA....15.1095L}; the other one containing chemical abundances is by a parallel convolutional neural networks, SPCANet~\citep{2020ApJ...891...23W}. 
LASP derived the atmospheric parameters (\teff, \logg, [Fe/H]) based on spectral-fitting the interpolation of the empirical spectra set--ELODIE v3.1~\citep{2007astro.ph..3658P}. While it is incapable of measuring chemical abundances. Several non-physical truncations appear in the density distribution of the LASP results (in the left-top panel of Fig.\ref{figure4}) due to it using four interpolators of templates in different temperature ranges. 
The stellar parameters and chemical abundances by SPCANet~\citep{2020ApJ...891...23W} are transferred from APOGEE-ASPCAP~\citep{2016AJ....151..144G} catalog based on the data-driven method. Independent and atmospheric model-based chemical abundance measurements are essential for LAMOST MRS, not only for the examination of the previous catalog but also to add physical support for each element.

In this paper, we aim to analyze the medium-resolution spectra of LAMOST DR8 MRS by domain-adaptation fitting the MARCS atmospheric model theoretical synthesis spectra. Five stellar parameters (\teff, \logg, [Fe/H], $v_{mic}$ and $v_{mac}$) and eleven chemical elements (C, Na, Mg, Si, Ca, Ti, Cr, Mn, Co, Ni and Cu) of LAMOST MRS are firstly derived based on the atmospheric model by \csnn. The main structure of this paper is organized as follows. Section.\ref{sec2} describes the observations of LAMOST DR8 MRS and the theoretical spectra generated. Section.\ref{sec3} introduces the method used to match the observed spectra and the synthetic model spectra. Section.\ref{sec4} shows the results of the stellar parameters and chemical abundances of LAMOST DR8 MRS, also the validation and analysis of the results are displayed. The final section is the summary.

\section{LAMOST-\Rmnum{2} DR8 MRS} 
\label{sec2}
\subsection{Observations and Data Reduction}

The Large Sky Area Multi-Object Fiber Spectroscopic Telescope (LAMOST) is a 4-meter reflecting Schmidt telescope located at the Xinglong Observatory (40.4174$^{\circ}$ N, 117.5006$^{\circ}$ E) in Hebei, China. It has been operating normally for 10 years. The Medium-Resolution Spectroscopic survey has been carried out for 5 years so far, and its detection limit is G$ \sim $17 mag. The light of sources passes through 4000 fibers and is dispersed with a resolution of 7500 into two separate bands (blue arm: 4950-5350 \AA~and red arm: 6300-6800 \AA), and recorded by 32 charge-coupled devices corresponding to 16 spectrographs. Each non-time-domain spectrum is combined using multi-exposes (at least three exposures) spectra for a star. The Tu-Ar arc lamp spectra exposed three times during one observing night are used in the process of wavelength calibration of MRS spectra. Unfortunately, no flux calibration process is done in MRS data reduction, because there are not enough suitable flux standard stars for each exposed plate (private communication with the LAMOST team).

\begin{figure}[htbp]
\centering
\includegraphics[width=0.5\textwidth]{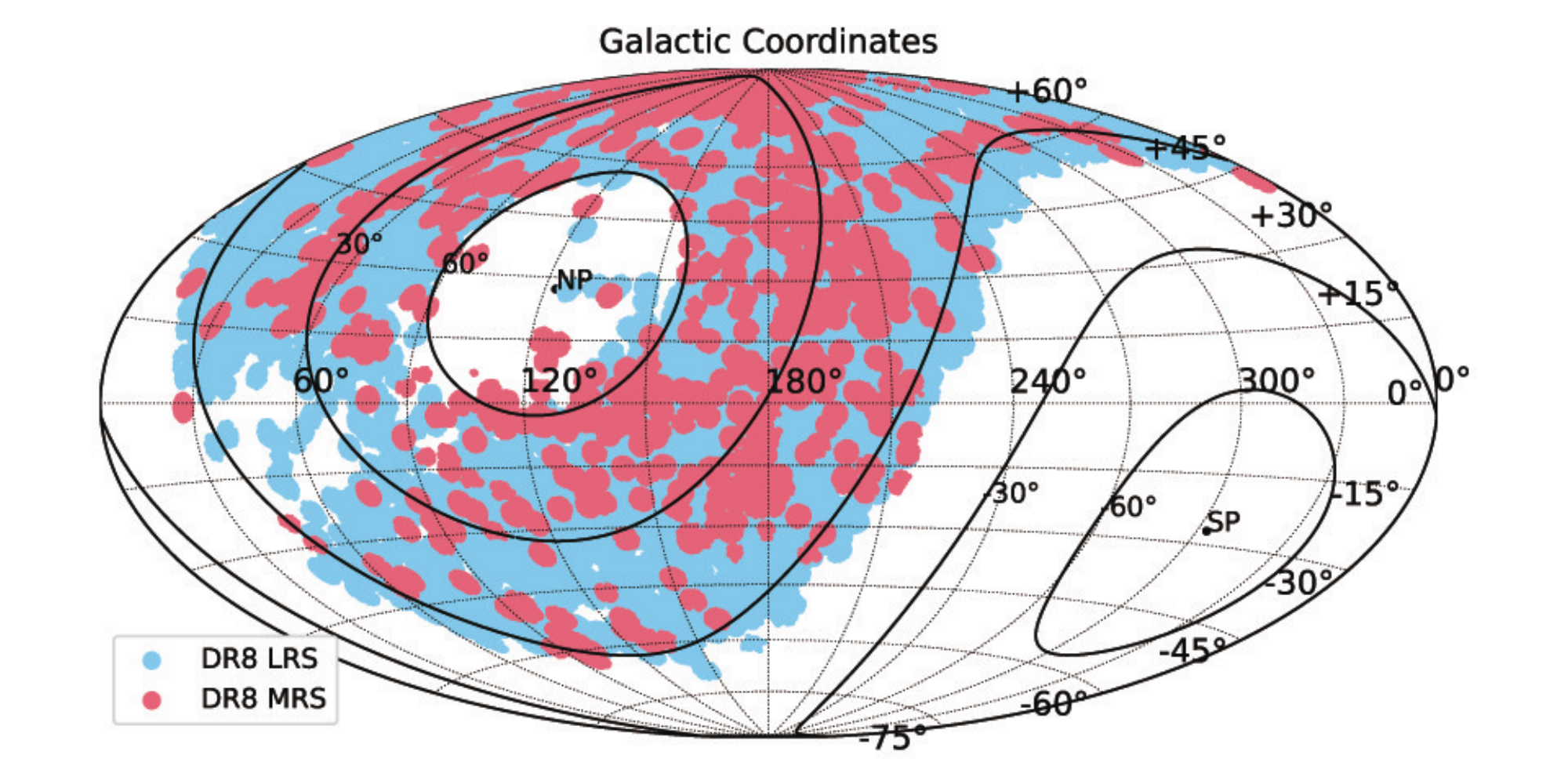}
\caption{LAMOST DR8 observations' footprints. The low-resolution survey (LRS) is shown in blue and the median-resolution survey (MRS) is in red.}
\label{figure1}
\end{figure}

\begin{figure}[htbp]
\includegraphics[width=0.47\textwidth]{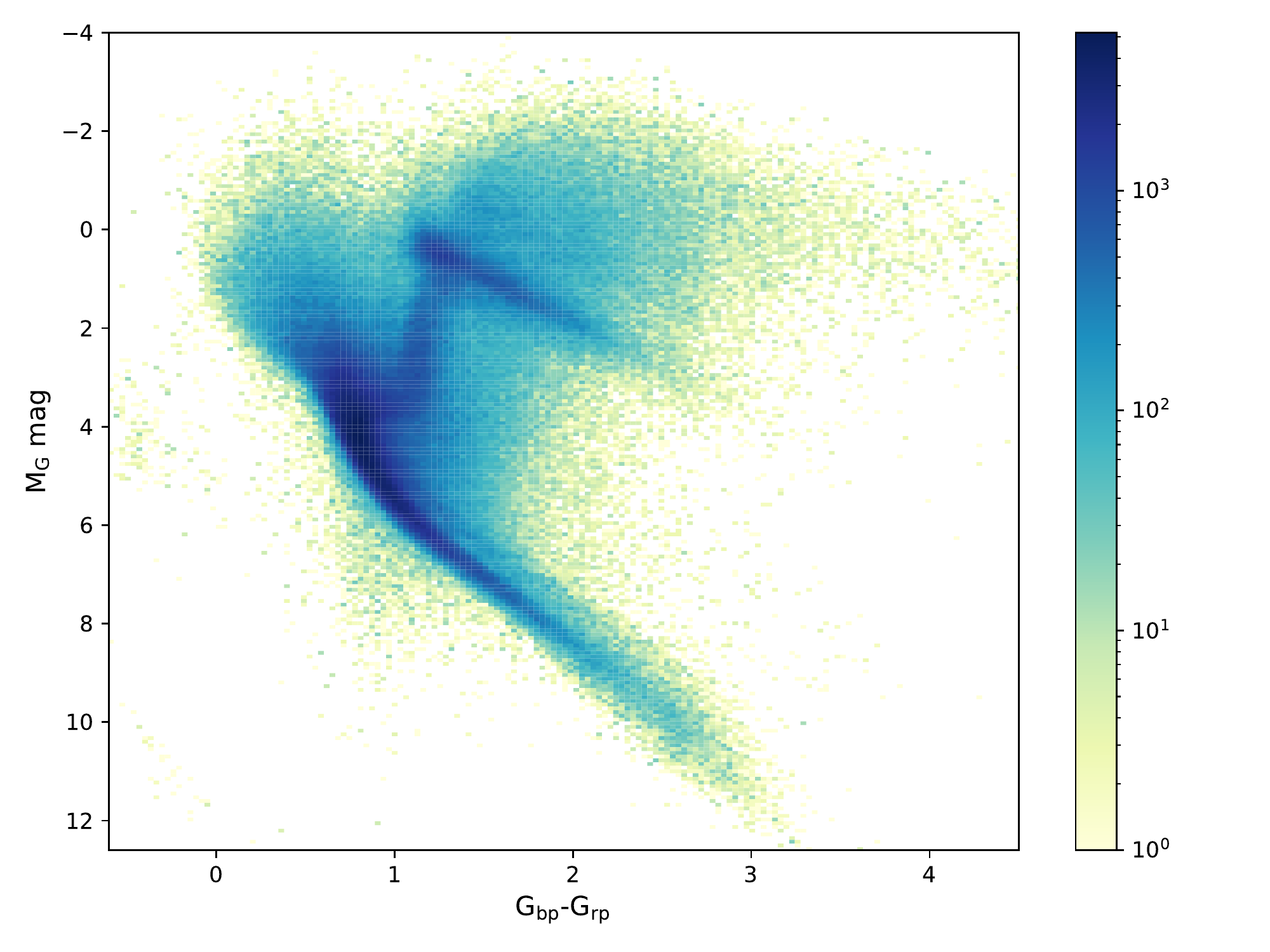}
\caption{Color-absolute magnitude diagram of the LAMOST DR8 MRS objects color-coded by number density in log-scale. Here, the magnitude is calculated without extinction correction.}
\label{figure2}
\end{figure}

\begin{figure*}[!htb]
\includegraphics[width=0.95\textwidth, ]{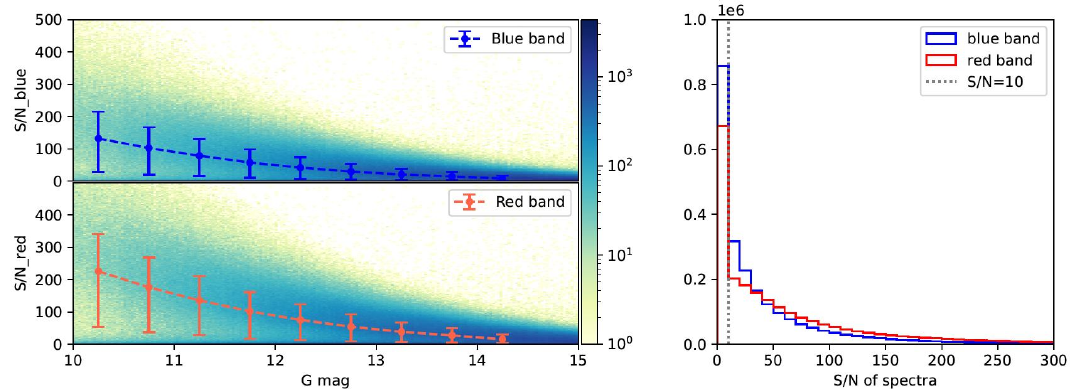}
\caption{S/N density distributions of the LAMOST DR8 MRS blue (top panel) and red (bottom panel) arms as a function of the Gaia photometric G magnitude are shown above. the errorbar in red represent S/N of blue (top panel) and red arms (bottom panel) in different G mag bins, each of which spans 0.5 magnitude.}
\label{figure3}
\end{figure*}

LAMOST 8th data release (DR8) published 1,479,127 non-time-domain spectra and 4,559,091 time-domain spectra. The footprints of LAMOST DR8 observations (both low-resolution and medium-resolution survey) and the color-absolute magnitude (without extinction correction) diagram of the sources are shown in Fig.~\ref{figure1} and Fig.~\ref{figure2}, respectively. The distributions of the spectral signal-to-noise (S/N) of LAMOST DR8 MRS blue and red arm spectra as a function of the $Gaia$ photometric $G$ magnitude employed from $Gaia$ DR3 are shown in Fig.~\ref{figure3}.~\citet{2017ApJ...843...32T,2020RAA....20...51Z} used theoretical spectra to explore the variation of star labels accuracy with S/N and found that the stellar labels suffering large biases for the "mock" medium-resolution (R=6000/7500) spectra with S/N lower than 10. Most of the LAMOST MRS spectra with low data quality as S/N lower than 10 are considered not suitable for deriving stellar parameters or chemical abundances using spectral fitting methods. So, we excluded these low-S/N ($ \le $ 10) spectra from our dataset. The radial velocities of LAMOST MRS spectra are derived by cross-correlating with ~2000 Kurucz model templates and calibrated by the RV standard stars, and the precision of the MRS RVs reaches 1.36 $km s^-1$ with signal-to-noise (S/N) higher than 10~\citep{2019ApJS..244...27W}. For this work, the relative radial velocities among multiple exposure spectra of one observation were calculated by self-correlation and corrected during the co-adding process. We separately pseudo-continuum normalized the blue and red arms of LAMOST MRS spectra by fitting a 5-order polynomial through multiple iterations and then Doppler-shifted them to the rest-frame wavelength with their radial velocities respectively. To get reliable stellar parameters, we need to recognize and exclude the spectroscopic binary or multi-system spectra. Here, we eye-recognized a sample of 500 spectra with obvious double or triple lines and a sample of 1000 spectra of single stars to train a convolutional neural network (CNN) model to classify the spectroscopic multi-system spectra. The CNN model can identify target spectra with a test accuracy of 96\%. The spectroscopic binary spectra with double line features will be processed by a light curve-based binary spectral model to measure their stellar parameters~(Chen et al. accepted).

\subsection{Stellar Parameters provided by LASP and SPCANet}

\begin{figure*}[htp]
\centering
\includegraphics[width=0.95\textwidth]{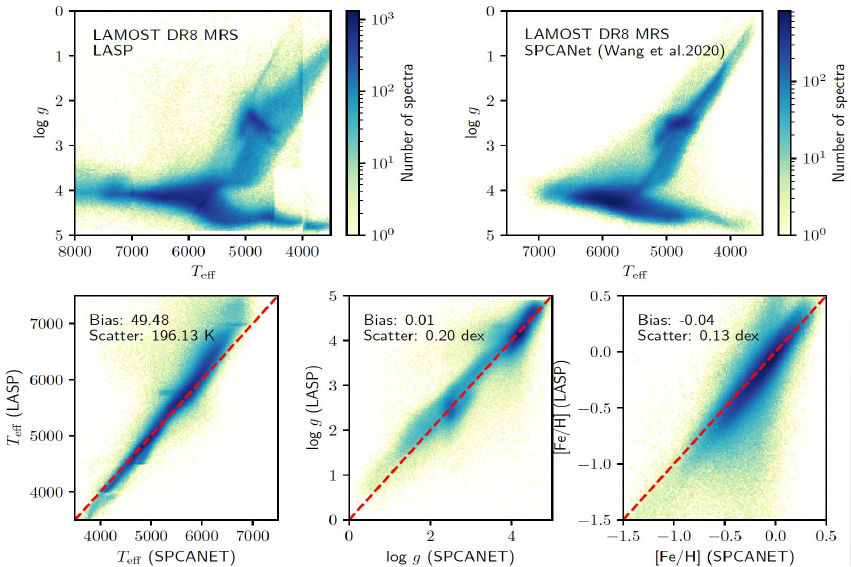}
\caption{Density distributions in the Kiel-diagram of LAMOST MRS stellar parameters by LASP(left-top) and SPCANet~\citep{2020ApJ...891...23W}(right-top). Comparisons of three stellar parameters between the two methods are shown in the bottom three panels and the red dotted lines indicate the one-to-one lines.}
\label{figure4}
\end{figure*}

LAMOST Stellar Pipeline~\citep[LASP;][]{2015RAA....15.1095L} plays a continuous role in deriving basic stellar parameters of LAMOST MRS spectra based on ULySS package~\citep{2009A&A...501.1269K}, fitting with the empirical spectral templates--ELODIE v3~\citep{2007astro.ph..3658P}. Three basic stellar parameters of 51.32 \% non-time-domain and 11 \% time-domain spectra were officially provided. Besides, a catalog including 3 basic stellar parameters and 13 chemical abundances are reported employing an updated version of neural networks SPCANet~\citep{2020ApJ...891...23W} by transferring APOGEE DR16 ASPCAP~\citep{2020AJ....160..120J} stellar parameters and chemical abundances to LAMOST MRS spectra. Fig.~\ref{figure4} shows the Kiel -diagram of LASP and SPCANet results of LAMOST DR8 MRS. It can be found that significant differences exist in the Kiel-diagram by the two different methods. Artificial structures appear in dwarfs at 7000 K, 4500 K, and 4000 K and giants at 4000 K in LASP results due to fitting different kinds of empirical interpolated spectra in different parameter spaces~\citep{2015RAA....15.1095L}. One-to-one comparisons of the stellar parameters also reflect that the two methods show a dispersion of 196 K and 0.2 dex in the effective temperature and surface gravity. The \teff~by LASP is overestimated for hot stars compared with SPCANet results. For [Fe/H], two kinds of results show good agreement for most stars except part of stars with [Fe/H]$_{SPCANET}$ in [-0.5, 0] dex having a slope greater than 1 component. It should be mentioned that the results of the label-transfer method suffer from the selection effect of the training labels from MRS-APOGEE common stars. And, the elemental abundances predicted by SPCANet are only available for the stars with effective temperatures between 4000 K and 5000 K, limited by the elements availability of the training samples.

\section{Methods}
\label{sec3}

In order to increase the physical interpretability and get rid of the disadvantages of the data-driven approach, we use a stellar atmospheric model-driven method to obtain the stellar parameters and chemical abundances of LAMOST DR8 MRS. The forward modeling idea is used to obtain an atmospheric model-based synthetic spectrum to model the observed spectrum, and further realize the transferring of the stellar parameters and chemical abundances from the synthetic model to observed spectra. However, due to the imperfection of the existing stellar atmospheric model, the actual observed conditions (such as skylight, anthropogenic light pollution), the instability of the instrument effects, and the errors introduced by the data processing, a gap exists between the synthetic spectra and the observed spectra~ (as Fig.~\ref{figure6} up-left panel shows). Traditionally, the spectral pixels with large flux differences between synthetic and observed spectra are masked and excluded to optimize the fitting process, but it would be effective only for the stars close to the standard stars which are used to evaluate the synthetic spectra.~\citet{2021ApJ...906..130O} proposed that the domain adaptation method can automatically solve this problem based on deep-learning methods.

\subsection{The theoretical synthetic spectra}

Here, we generate the theoretical synthesis spectra of FGKM type stars (\teff: 3500-8000 K) as the theoretical part of the training data sets. We use 1D LTE MARCS atmosphere models~\citep{2008A&A...486..951G} and the spectrum synthesis code Turbo-Spectrum for radiative transfer~\citep{2012ascl.soft05004P} to calculate the theoretical spectra. The solar chemical abundances of~\citet{2007SSRv..130..105G} are adopted here and the atomic and molecular line lists are from the latest version of Gaia-ESO survey line lists~\citep{2015PhyS...90e4010H,2021A&A...645A.106H}. It is worth mentioning that the line lists are state-of-art covering the visual bands but it is not perfect for cool stars with temperatures below 4300 K because of the incompleteness of molecular lines.

Unlike most of the official pipelines, we do not follow a fixed step to get theoretical spectra on parameter grid points but randomly sample in a convex hull of the MIST isochrones within an age range of 1-10 Gyr and a metal abundance range of [-1.5, +0.5] dex, under the 17-dimensional stellar labels (effective temperature \teff, surface gravity \logg, metal abundance [Fe/H], macroturbulence $v_{mac}$ and microturbulence $v_{mic}$ and 11 elements [X/Fe]). The purpose of introducing isochrones here is to limit the range of parameters for the generation of the theoretical spectra, which can cover most of the parameters space of LAMOST MRS observations, but of course, individual stars at particular evolutionary stages are lost. We generate 50,000 spectra with stochastic combinations of stellar labels within the constraint ranges above, which is a relatively time-consuming task. About 50,000 CPU hours are spent using a 16-computer distributed cluster armed with a SPARK computing system~\citep{Zaharia2016}.

\subsection{Reducing the Gap between LAMOST Observations and the theoretical synthesis spectra}

Whether a physical-driven method or a data-driven method, the essence is to synthesize a spectrum model to fit the observation. The assumption is that the spectra with the same morphology have the same stellar atmospheric parameters and chemical abundances, which is the idea of forward modeling (a data model or a physical model). When using spectral model-fitting technology to derive stellar parameters, we need to ensure that both the observed spectra and the model spectra follow the same distribution of the fluxes in the n-dim pixels space. Recent studies~\citep{2020MNRAS.498.3817B,2021ApJ...906..130O} have shown that the distributions of the theoretical and the observed spectra do not overlap in the high-dim fluxes space, so-called "the gap". The main reason is that the theoretical model suffers the defects of the internal assumptions of the physical model and the imperfect line lists, such as the determination of oscillator strength $log~\textrm{gf}$, which can't perfectly express the spectral features the observed spectra contained. To deal with these situations, most researchers adopt the approach by masking the "bad" pixels of which the difference between the observed and synthetic spectral fluxes suffer large extent according to the standard stars~\citep{2019ApJ...879...69T}. ~\citet{2020MNRAS.498.3817B} superimposed simulated Gaussian noise, rotational and radial velocities, masking disturbed regions, and bad pixels to augment the theoretical spectra to minimize the synthetic gap with the Gaia-ESO UVES spectra, which would lead to the loss of much spectral information. However, the \csn~\citep{2021ApJ...906..130O,2020arXiv200703112O} can handle a variety of observed effects and noise patterns instead of masking spectral regions and pixels.

The unsupervised domain adaptation method \csn~\citep{2021ApJ...906..130O,2020arXiv200703112O} has been shown to reduce the gap between the synthesis spectra and observed spectra well. The main idea of this method is to map the synthesis and observed spectra to a feature space through an auto-encoder and to separate the shared and split features of the synthesis and the observed spectra by using a Generative Adversarial Network~\citep[GAN;][]{2022arXiv220300667C}. Accordingly, the shared features represent the information shared by the theoretical synthesis spectra with observed spectra; in theory, these should represent the internal physical properties of the spectra. The split features mainly express the unique information that the observed domain contains, which could be the instrumental profiles (e.g., line spread function), errors due to the data reduction (e.g., pseudo-continuum normalization), or variations that are not found in the theoretical spectra (e.g. telluric lines from the Earth's atmosphere). Not only does Cycle-StarNet separate the shared and split features in the spectra, but it also accomplishes the task of transferring them to different domains through cross-combining features. For instance, we can generate a theoretical synthesis spectrum with the observed pattern of a given LAMOST MRS observed spectrum by combining the shared features of the theoretical synthesis spectrum and the split features of the LAMOST MRS spectrum, then generate the spectral fluxes by decoding the combined features. Conversely, we can achieve the transfer to the theoretical domain by stripping the split features from the observed spectra.

To achieve the purpose above, the \csn is designed with 3 main components: an auto-encode network of the theoretical domain to map the synthesis spectra to the feature space, an auto-encode network of the observed domain to map the observed spectra to the feature space, and a generative adversarial network to separate the share and split feature space by judging the cycle- or transfer- generative spectrum. We modified the structure and hyper-parameters (such as convolution kernel sizes, and the feature space size) in the original version of \csn according to the bands and resolution characteristics of LAMOST MRS, making it adaptable to domain-transfer learning between LAMOST MRS observed and MARCS atmospheric model synthetic spectra.

\subsection{\csn training and test}

To construct the training data set, we first build a theoretical spectrum generative model~\citep{2019ApJ...879...69T} based on the theoretical synthesis spectra, which can synthesize theoretical synthesis spectra with any combination of stellar parameters within a certain range. Then we randomly select 100,000 LAMOST MRS observed spectra and synthesize 100,000 theoretical spectra. Each of the spectra sets is divided into three parts: 80,000 samples for training, 10,000 for validation, and 10,000 for testing \csnn. It is worth mentioning that the number of theoretical spectra in the training samples is the same as that of the observed ones, but they are not paired.

\begin{figure}[htbp]
\centering
\includegraphics[width=0.5\textwidth]{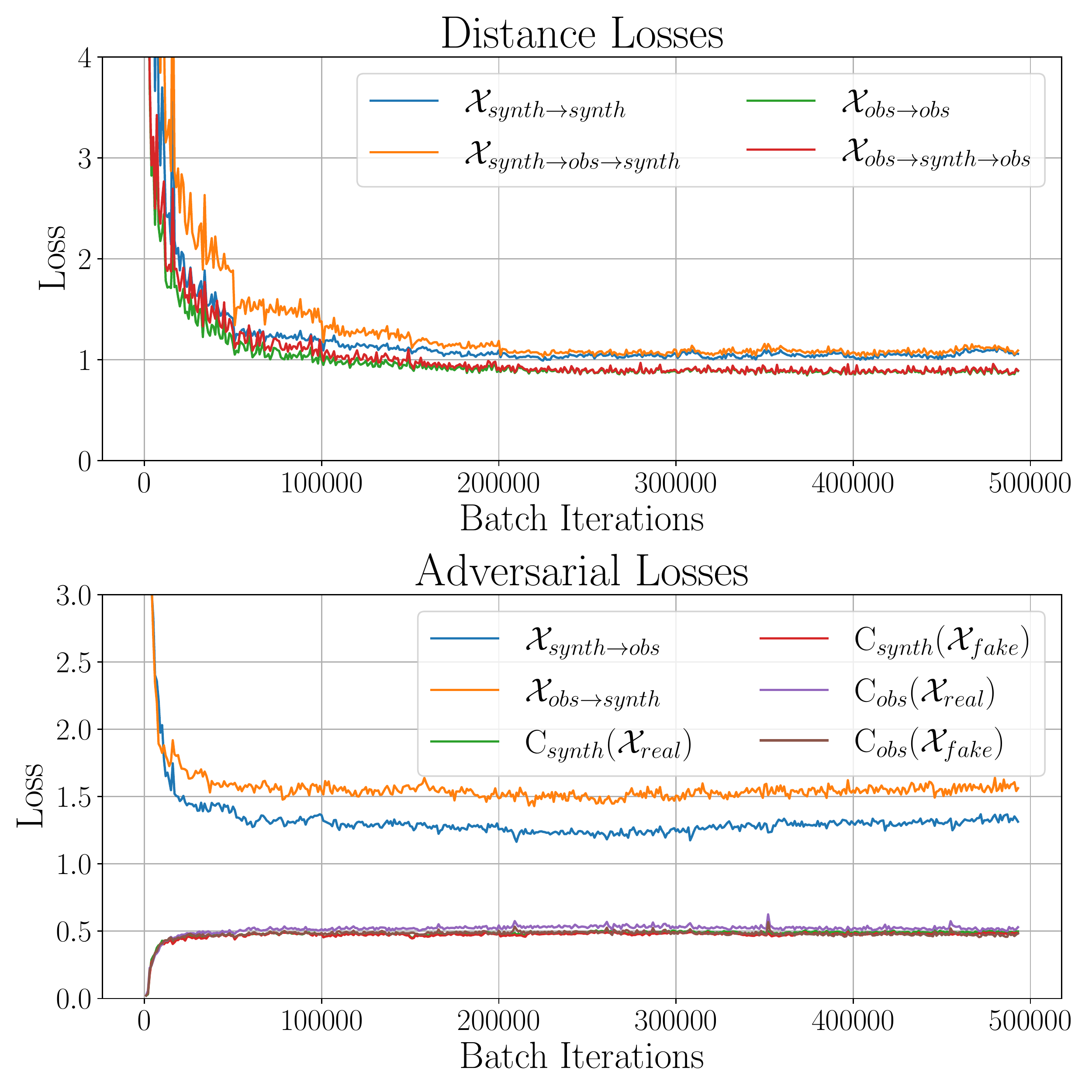}
\caption{Training losses of the \csn as a function of epochs. The loss function of \csn consist of three parts: transfer-loss ($X_{synth->obs}$, $X_{obs->synth}$), cycle-loss ($X_{synth->synth}$, $X_{obs->obs}$, $X_{synth->obs->synth}$, $X_{obs->synth->obs}$) and adversarial-loss($C_{synth}(X_{real})$, $C_{synth}(X_{fake})$, $C_{obs}(X_{real})$, $X_{obs}(X_{fake})$).}
\label{figure5}
\end{figure}

We trained \csn on 500,000 iterations to optimize the weight parameters of its neuron nodes by minimizing the loss function (as shown in Fig.~\ref{figure5}). The loss function of \csn is weighted-composed of three parts: transfer-loss, cycle-loss, and adversarial-loss. The transfer-loss constrains the chi-square distance between the theoretical synthetic spectrum and the "synthetic" spectrum transferred by the observed spectrum, and the chi-square distance between the observed spectrum and the "observed" spectrum transferred by the theoretical synthetic spectrum; the cycle-loss describes the distances between the spectrum, both the theoretical and the observed, and the spectrum after two times transfers back to their original domains; the adversarial loss allows the spectra to achieve unsupervised domain transfer in different domains. About 10 hours was cost to train the model each time using an NVIDIA Tesla v100 GPU.

\begin{figure}[htbp]
\centering
\includegraphics[width=0.47\textwidth]{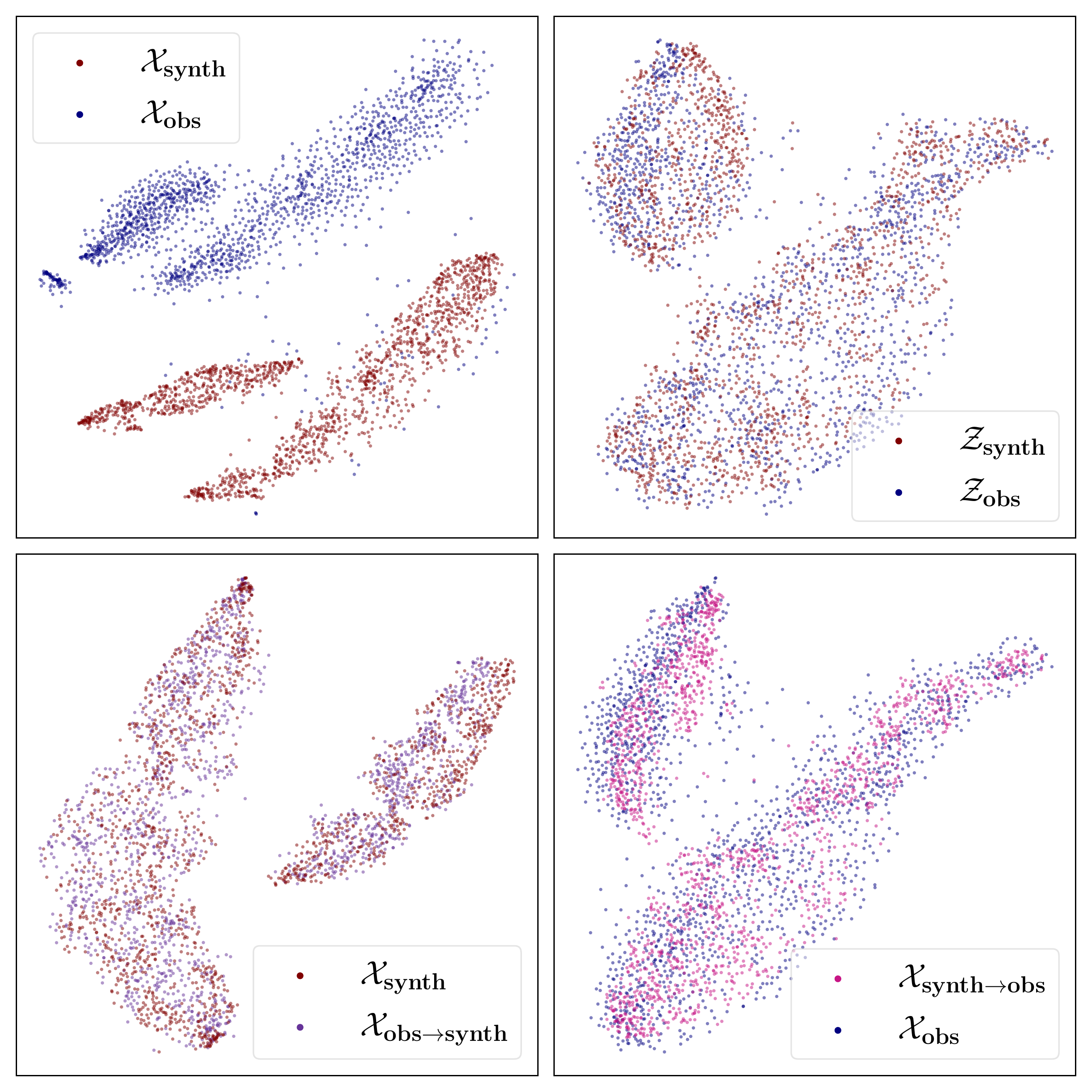}
\caption{T-SNE test pairs of two of 10,000 synthetic spectra $X\_{synth}$, 10,000 LAMOST MRS observed spectra $X\_{obs}$, 10,000 $X\_{synth\_obs}$ spectra, 10,000 $X_{obs\_synth}$ in the observed domain (bottom-right panel), synthetic domain (bottom-left panel) and their shared feature (top-right panel) mapped into two-dim space by \csnn.}
\label{figure6}
\end{figure}

We used 10000 LAMOST MRS spectra and 10000 theoretical synthetic spectra to test the model. A method, t-distributed stochastic neighbor embedding (t-SNE), is employed to examine the distributions of the spectral sets before and after domain transfer, also their features in hidden space (shown in ~Fig.\ref{figure6}). T-SNE is a statistical method for visualizing high-dimensional data by a non-linear reduction in a low-dimensional space of two or three dimensions. 4000-dimensional flux spectra can be modeled as two-dimensional points in the way that similar spectra are modeled by nearby points and dissimilar objects are modeled by distant points with high probability. 

The T-SNE test results show that there is a clear gap between the LAMOST and synthetic spectra, which is why it is inappropriate to match them directly. The spectra transferred by the \csn model make their distributions in the same domain overlap together, and the shared features of the theoretical and observed spectra in the hidden space follow the same distribution, which indicates that the extracted features of the theoretical and observed spectra during the transfer process are physical-based, not data-based. The \csn successfully morphs the synthetic domain to the observed domain, and effectively reduces the fitting residuals, especially for the cool stars.

\subsection{Domain-transfer Spectral Fitting}

After finishing the model training, we can search the best-fit synthetic spectrum for a given observed spectrum either in the synthetic domain (SD) or the observed domain (OD). The chi-square distance between the original spectrum and the domain-transferred spectrum is minimized (according to equations 10 and 11 of the paper~\citet{2021ApJ...906..130O}) and the corresponding stellar labels are obtained for the observed spectrum. For LAMOST DR8 MRS spectra, we process the domain-transfer fitting in both the synthetic and observed domain by searching the best-fit stellar labels based on the L-BFGS algorithm which is an iterative method to solve unconstrained nonlinear optimization problems. The minimized fitting distance in the synthetic domain is obviously smaller than the one in the observed domain because transferring the observed spectrum to the theoretical domain is a process of de-noising while transferring the theoretical spectrum to the observed domain is the opposite. The whole process is accomplished under the GPU framework and costs about one week.

\subsection{Constrain of the surface gravity with the photometric and astrometric data}

The region of LAMOST MRS spectra is not very sensitive to surface gravity~\citep{2020RAA....20...51Z,2020ApJ...891...23W}. So it is difficult to obtain accurate \logg~only using spectral information. Fortunately, the Gaia EDR3 provides us with accurate parallaxes for most of LAMOST MRS stars. Combined with relatively complete photometric data, the surface gravity of LAMOST objects can be further constrained as follows.

In this work, we employee the Gaia distances ~\citep{2021AJ....161..147B} and Ks magnitudes of 2MASS survey~\citep{2006AJ....131.1163S}. 
The Milky Way 3D dust map ~\citep{2019ApJ...887...93G} is used to correct the extinction for the apparent magnitude. 
After deriving the absolute magnitude of Ks ($M_{Ks}$) by 
$$ 
M_{Ks} = K_{s} - 5*log (\frac{Distance}{10}) - A(K_{s}), 
$$ 
we calculate the bolometric luminosity of a star by
$$   
log(\frac{L_\textrm{bol}}{L_{\textrm{bol},\odot}}) = -0.4*(M_{Ks} + BC_{Ks} - M_{bol, \odot}),
$$
where, $BC_{Ks}$ is the bolometric correction based on \citet{2016ApJ...823..102C}, $M_{bol, \odot}$ is the bolometric absolute magnitude of Sun (we set as 4.7554).
Then, the surface gravity can be literally calculated as GALAH DR3~\citep{2021MNRAS.506..150B} by 
$$ 
log~g = log~g_{\odot} + log(\frac{M}{M_{\odot}}) + 4*log(\frac{\textit{T}_\textrm{eff}}{\textit{T}_{\textrm{eff},\odot}}) - log(\frac{L_\textrm{bol}}{L_{\textrm{bol},\odot}}), 
$$
 where $M$ is the mass of star, $M_{\odot}$ is the mass of Sun, $\it{T}_\textrm{eff}$ is derived from LAMOST MRS spectrum, $\textit{T}_{\textrm{eff},\odot}$ is 5777 K, $log~g_{\odot}$ is 4.44 dex. Here, the stellar mass $M$ is the only unknown quantity, we estimate the stellar mass by interpolating MIST isochrones as given: spectra-derived \teff, \logg, [Fe/H], and distance from~\citet{2021AJ....161..147B}. Then we can further calculate the new \logg~by the above 3 equations, and then update the stellar mass by interpolating isochrones with the updated \logg, iterate the steps above until the stellar mass converges, and finally get the star mass $M$.

\section{Results}
\label{sec4}

\begin{figure*}[htbp]
\centering
\includegraphics[width=0.95\textwidth]{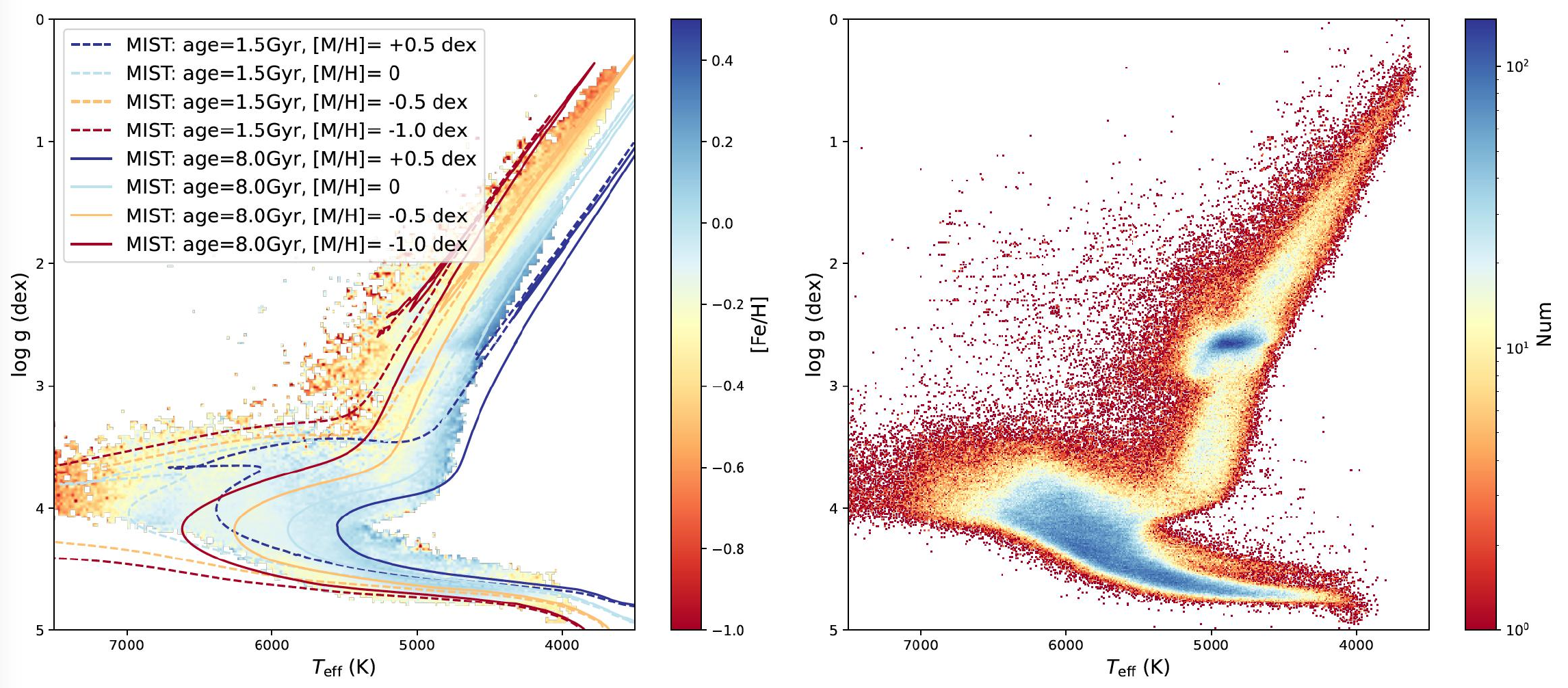}
\caption{Distribution of LAMOST DR8 MRS stellar parameters (Flag$_{quality}$=0) by \csn in Kiel diagram color-coded by [Fe/H] (left panel) and number density (right panel). The MIST isochrones are superimposed on the left diagram. The dashed and solid lines represent ages of 1.5 Gyr and 8 Gyr, respectively. For each age, the color red, yellow, light blue, and blue represent [Fe/H] of -1.0 dex, -0.5 dex, 0, and +0.5 dex, respectively.}
\label{figure7}
\end{figure*}

\begin{figure*}[htbp]
\centering
\includegraphics[width=0.95\textwidth]{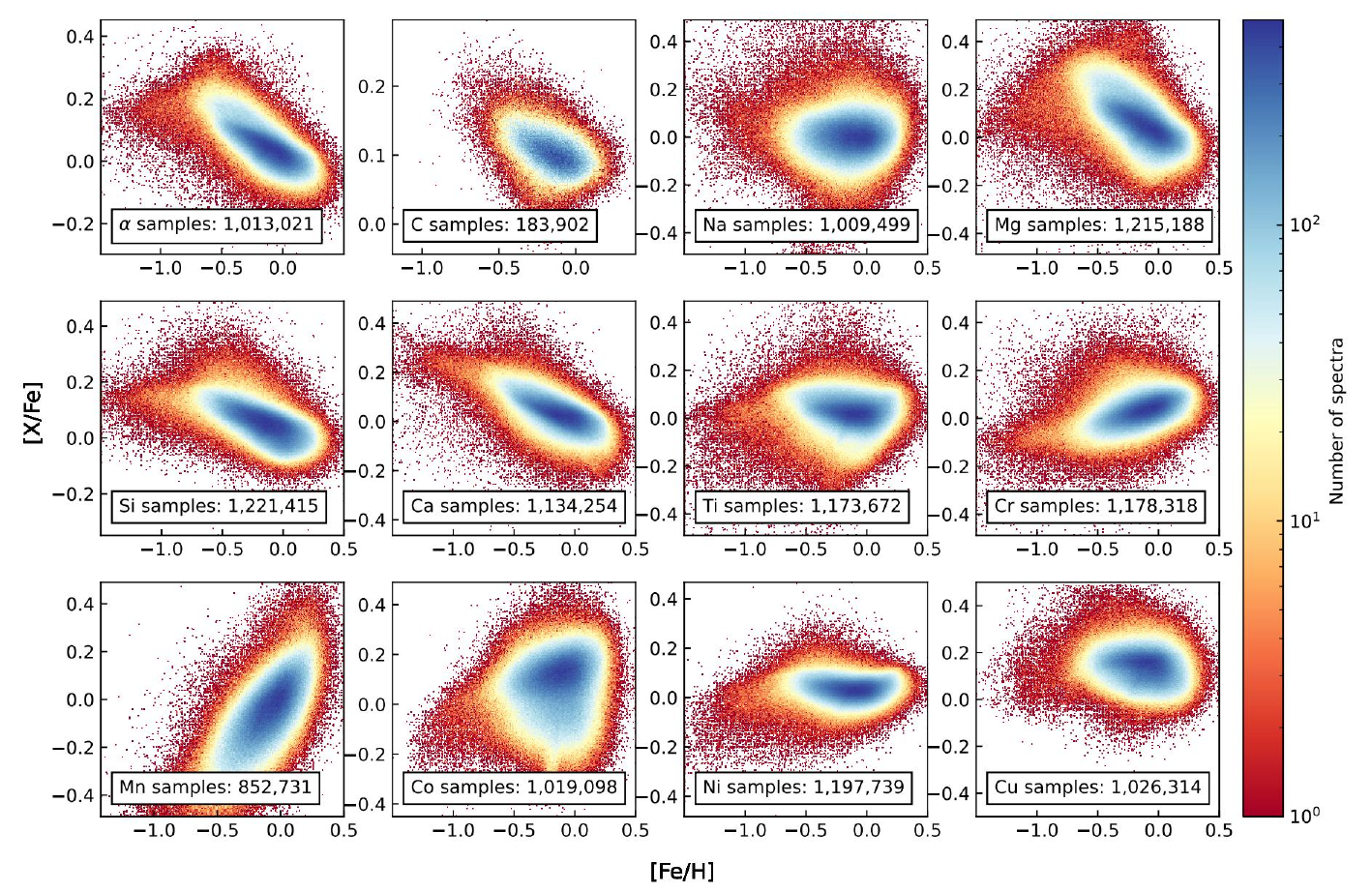}
\caption{Distribution of elemental abundances [X/Fe] estimated by \csnn, plotting as a function of [Fe/H] for LAMOST DR8 MRS, color-coded by number density in log-scale.}
\label{figure8}
\end{figure*}

We derived 5 atmospheric parameters (\teff,~\logg, $[Fe/H]$, $v_{mac}$, $v_{mic}$) and 11 chemical abundances of 1.38 million LAMOST DR8 MRS spectra with S/N $\geq$ 10. For 64.37\% of the spectra, their surface gravities are revised based on the photometric and astrometric information. The distribution of the Kiel-diagram color-coded by the metallicity [Fe/H] and number density is shown in Fig.~\ref{figure7}. Our results show good agreement with the MIST isochrones, and the main sequence and red clumps are evident. The temperature range of our results is from 3230 K to 8000 K. For stars with temperatures higher than 7500 K and lower than 3600 K, the stellar parameters become less reliable because of large inconsistency between the observed spectrum and the theoretical synthetic spectrum in this parameter space, which is reflected by the fitting $\chi^{2}$ value exceeds 2.5 and the large fitting error ($<$ 0.1) of \logg. Because these stars with few spectral features may be reconsidered as metal-poor stars by the code. It should be cautious when using the results of these stars hotter higher than 7500 K. 

We derive 11 elemental abundances, including C, Na, 3 alpha elements (Mg, Si, Ca), and 6 iron-peak elements (Ti, Cr, Mn, Co, Ni, Cu). Although in the synthesis process of the theoretical spectra we introduced carbon, nitrogen, and oxygen elements, limited by the resolving power of 7500 and the narrow band region of LAMOST MRS spectra, it is challenging to obtain accurate abundances of these three elements, particularly nitrogen and oxygen. We only keep carbon in the final catalog by constraining its fitting ${\chi}^2$ in a small range (${\chi}^2 \leq 1.5 $). The chemical abundances [X/Fe] as functions of [Fe/H] are shown in Fig.~\ref{figure8}, where \afe~is the mean of alpha elements. Through the~\afe~abundance distribution, we can see that the structures of the chemical thin disk, thick disk, and halo stars are revealed and the abundances seem to be underestimated for thick disk and halo stars. For iron-peak elements, most of them show similar distributions to previous studies, except that manganese has a larger slope and cobalt has a larger dispersion, compared to these studies based on different techniques and particularly spectra that are of much higher quality~\citep{2010A&A...522A..32R,2015A&A...577A...9B,2018A&A...616A..18E}. Non-LTE correction for Mn is necessary~\citep{2008A&A...492..823B}, while the theoretical model we used is the 1D LTE model.

\subsection{Accuracy Analysis}

Accuracy refers to the degree of closeness of the measurements to the "true" values of the quantity being measured. The $Gaia$ benchmark stars~\citep{2014A&A...564A.133J,2014A&A...566A..98B,2015A&A...582A..49H,2015A&A...582A..81J,2016A&A...592A..70H} can be used as anchor points for the "true" values, which show good consistency. Regrettably, the LAMOST MRS observations were unable to include Gaia benchmark stars due to their brightness limitations, and some of these stars are located in the southern sky, beyond LAMOST's observable regions. The PASTEL catalog~\citep{2010A&A...515A.111S,2016A&A...591A.118S} is a common option, which is a comprehensive and regularly updated database of stellar parameters, such as effective temperature, surface gravity, and metallicity. These stellar properties are compiled from a wide variety of sources, including published literature and other catalogs. We employ the PASTEL catalog to evaluate the accuracy of our stellar parameters in the following sub-section.\ref{subsub4-1}. Besides, the asteroseismic sample is one of the best choices to evaluate the accuracy of \logg, here we also choose the Kepler DR25 golden sample~\citep{2017ApJS..229...30M} and APOKASC catalog~\citep{2014ApJS..215...19P,2018ApJS..239...32P} to analysis our \logg~in sub-section.\ref{subsub4-2}. In sub-section.\ref{subsub4-3}, comparison with sources observed by other surveys, covering different bands with medium/high resolving power, can also reflect the accuracy among them.

\subsubsection{Accuracy evaluated by PASTEL catalog}
\label{subsub4-1}
We cross-match our results with the PASTEL catalog~\citep{2010A&A...515A.111S,2016A&A...591A.118S} getting 934 common stars. We excluded 82 suspect binary or multi-system objects by checking whether two or more sources appear in 5 arcsec view of their Pan-Starrs or 2MASS photometric images. After subtracting the null values of the parameters in the PASTEL catalog, the remaining available sample is 670, 244, 502 for \teff, \logg~and [Fe/H], respectively. We used the standard deviation of the weighted bias to measure the accuracy of the \teff, calculated as follows:

\begin{align*}
& \sigma_{X,i} = \sqrt{(\sigma_{X_{i,LA}})^2 + (\sigma_{X_{i,PA}})^2}, \\
& w_{X,i} = \frac{1}{\sigma_{X,i}}, \\
& \mu_{X,weighted} = \sum_{i}^{} \frac{(X_{i,LA} - X_{i,PA}) * w_{X,i}}{\sum w_{X,i}}, \\
& S_{X,weighted} = \sqrt{\frac{1}{N} \sum_{i}^{} (X_{i,LA} - X_{i,PA} - \mu_{X,weighted})^2},
\end{align*}
where $X$ represent one of \teff, \logg, [Fe/H], $\sigma_{X_{i,LA}}$ is error of LAMOST, $\sigma_{X_{i,PA}}$ is error of PASTEL, $w_{X,i}$ is the weight of star i, $\mu_{X,weighted}$ is weighted bias of stellar parameters and $S_{X,weighted}$ is the standard deviation of the weighted bias.

\begin{figure*}[htbp]
\centering
\includegraphics[width=0.95\textwidth, angle=0]{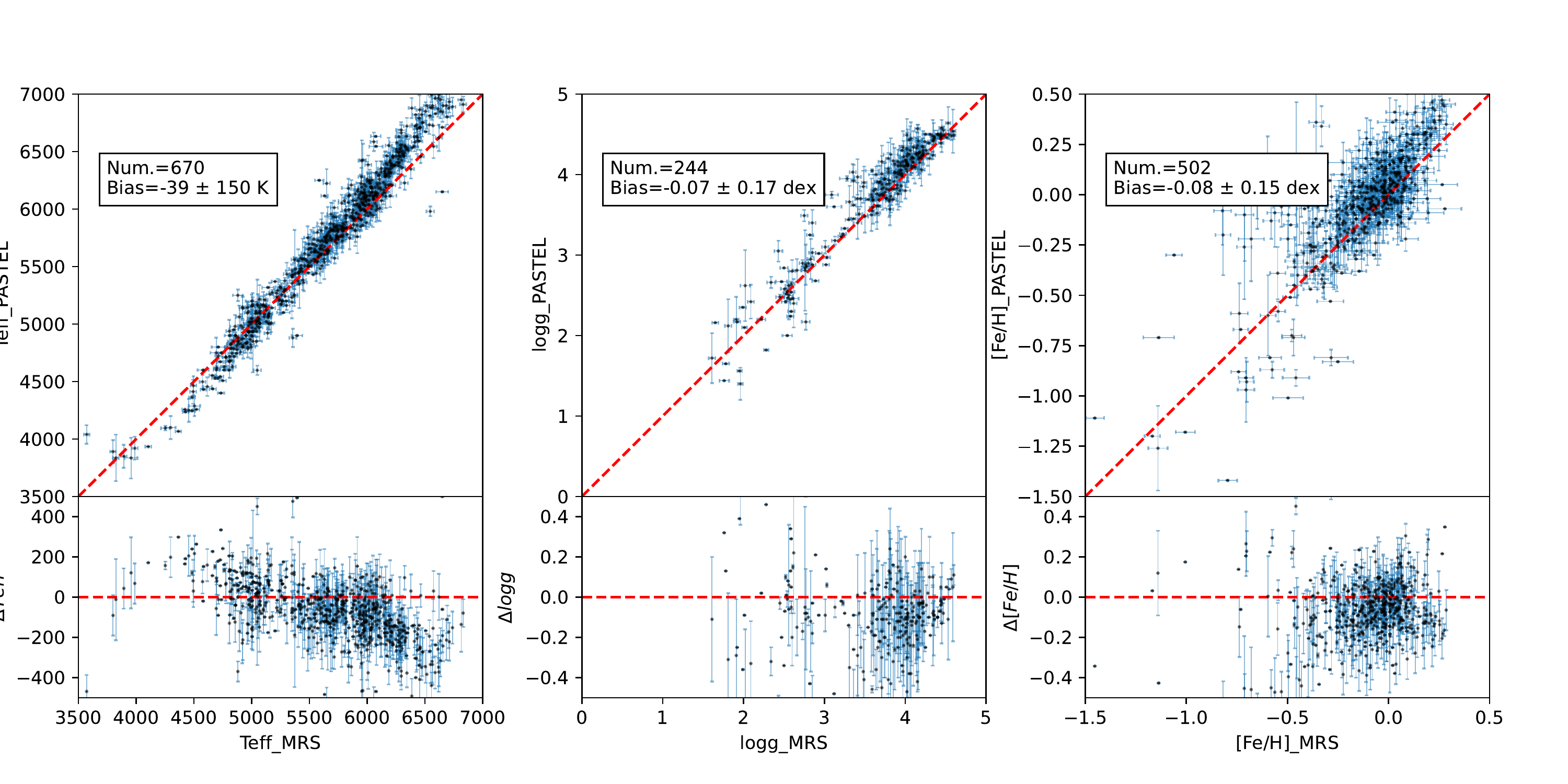}
\caption{Comparisons (top panels) and residuals (bottom panels) of three stellar parameters (\teff~in the left, \logg~in the middle, and [Fe/H] in the right) between LAMOST MRS by Cycle-StarNet and PASTEL catalog. Biases and the corresponding scatters of three are error-weighted and calculated.}
\label{Figure9}
\end{figure*}

Fig.\ref{Figure9} shows the biases with standard deviations of the differences between our results and the PASTEL catalog. It can be found that our \teff~exists a bias of -39 $\pm$ 150 K to PASTEL's measurements, 0.07 $\pm$ 0.17 dex for \logg, and -0.08 $\pm$ 0.15 dex for [Fe/H]. For \logg, we used other reference samples to evaluate, although its accuracy relative to PASTEL is shown in the figure. Here, we could only perform a statistical analysis of the accuracy of the overall sample rather than the accuracy of each parameter at different S/N levels, limited by the number of common stars we have.

\subsubsection{Accuracy evaluated by the asteroseismic sample}
\label{subsub4-2}
 
\begin{figure*}[htbp]
\centering
\includegraphics[width=1.0\textwidth, angle=0]{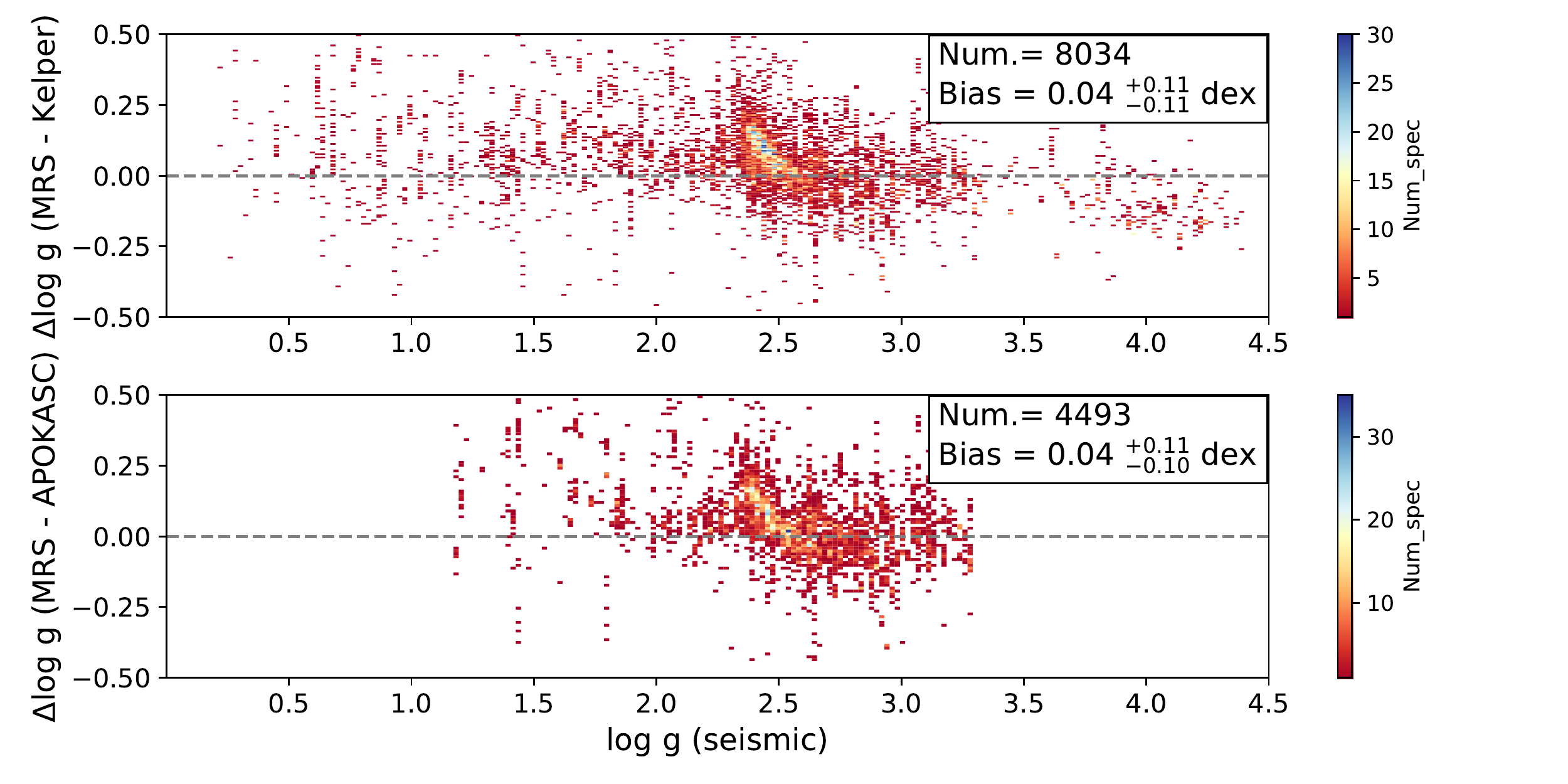}
\caption{Residuals density distribution of \logg~between LAMOST MRS by Cycle-StarNet and Kepler DR25 asteroseismic catalog in the top panel. The bottom panel show that of LAMOST MRS and APOKASC catalog for their common stars. In each panel, the value of the horizontal lines is 0. Bias is given as the median of the residuals combined with the 84\% quantile and 16\% quantile.}
\label{Figure10}
\end{figure*}

We employ two asteroseismic samples: one is Kepler DR25~\citep{2017ApJS..229...30M,2014ApJS..211....2H} in which the properties of golden samples contain \logg~are measured by asteroseismology; another one is APOKASC catalog~\citep{2014ApJS..215...19P,2018ApJS..239...32P} which report stellar parameters of 6676 evolved stars joint with APOGEE spectroscopic and Kepler asteroseismic data based on independent techniques. By cross-matching our sample with these two datasets, we get 8034 and 4493 common samples respectively. After comparing our results with the asteroseismic \logg~samples, we calculate the standard deviation of the weighted bias as accuracy, shown in Fig.~\ref{Figure10}. The bias between Kepler golden sample \logg~and ours is 0.04 dex with 84\% quantile of 0.11 dex and 16\% quantile of -0.11 dex. And for comparison with the APOKASC sample, we get nearly the same accuracy as the Kepler golden sample. We take 0.11 dex as the accuracy of our \logg. The proportion of giants (93\%:~\logg~$\leq 3.8$) in the MRS-Kepler golden sample is much larger than that of dwarfs (7\%:~\logg~$\ge 3.8$), so the accuracy here is more likely to reflect the properties of the giants, which also directly shown in the comparison of LAMOST MRS-APOKASC common samples. 

\subsubsection{Accuracy evaluated by other surveys}
\label{subsub4-3}

The development of large-scale spectroscopic surveys has provided us with many homogeneous spectral data sets. Their respective scientific goals determine their observed bands, resolving powers, and observation selection. These respective spectroscopic surveys inevitably adopt different spectral analysis methods to obtain the stellar atmospheric parameters and chemical compositions as they can. Here we employ data sets from three spectroscopic surveys (APOGEE DR16, GALAH DR3, RAVE DR6) to analyze the accuracy of LAMOST MRS stellar labels results.

\begin{figure*}[htbp]
\centering
\includegraphics[width=0.9\textwidth]{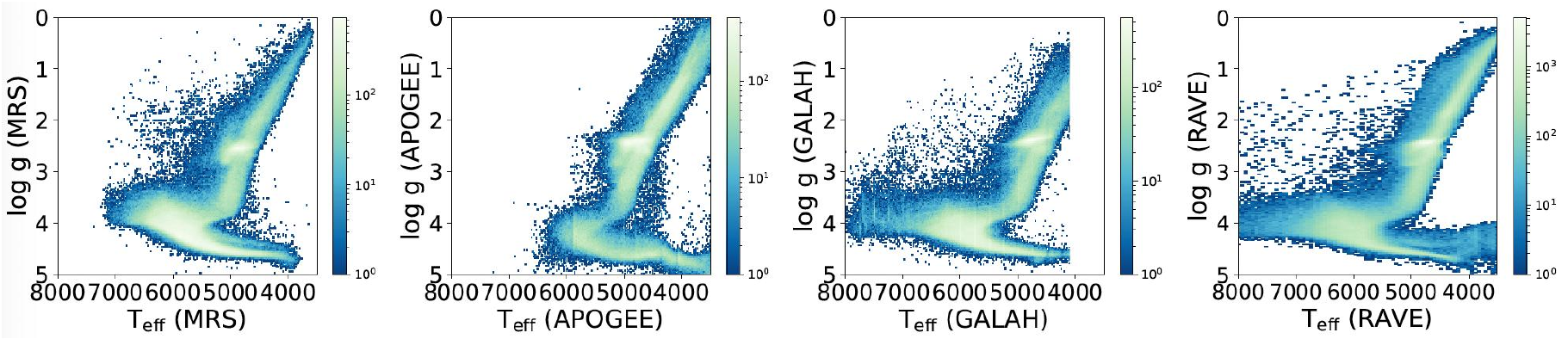}
\caption{Density distribution in Kiel-diagram of LAMOST DR8 MRS by \csnn, APOGEE (ASPCAP) DR16, GALAH DR3, and RAVE DR6. The color indicate number densities of spectra in log-scale.}
\label{figure11}
\end{figure*}

\begin{figure*}[htbp]
\centering
\includegraphics[width=0.9\textwidth]{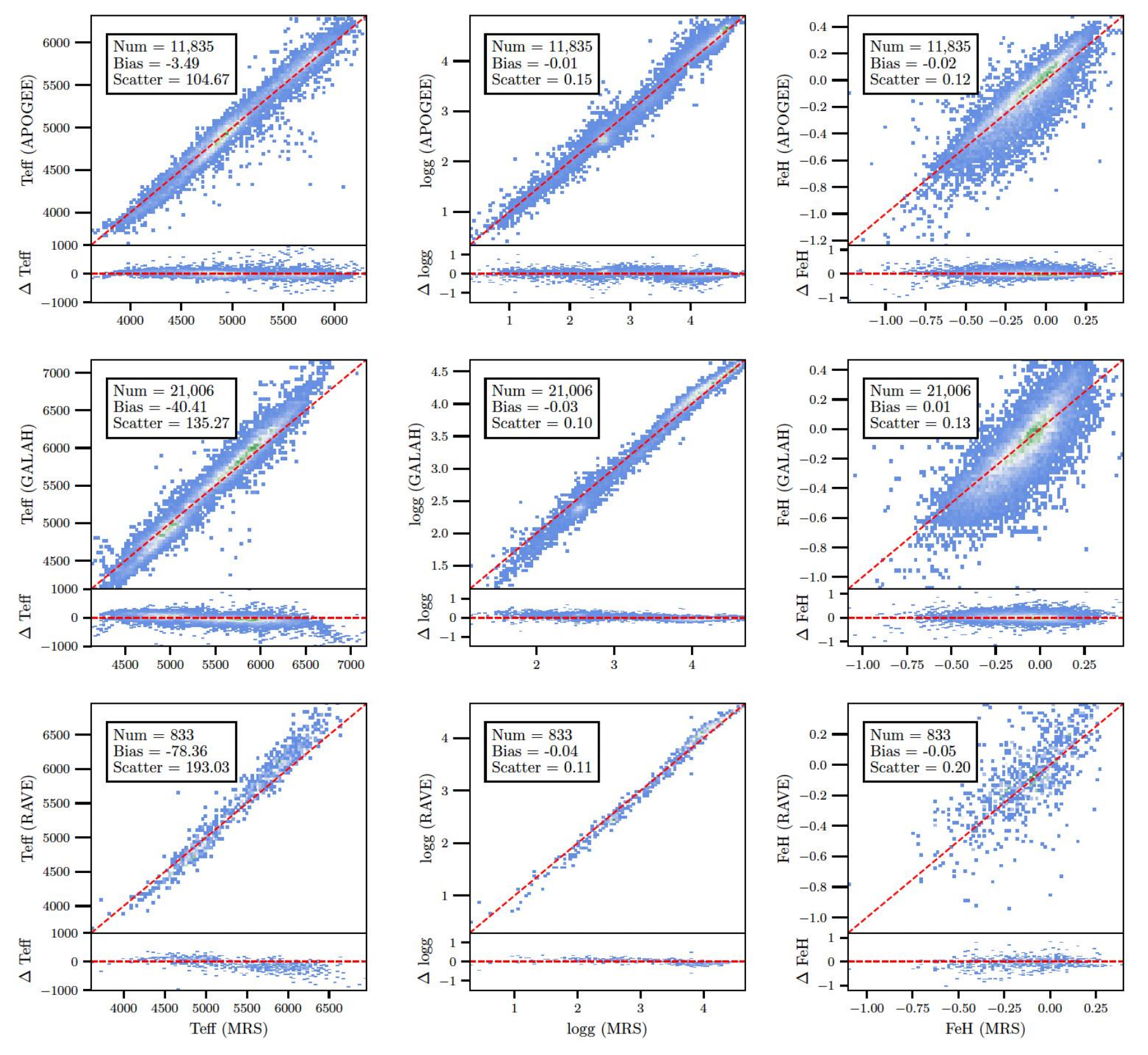}
\caption{Comparison of three stellar basic parameters between LAMOST DR8 MRS (\csnn) and APOGEE (ASPCAP) DR16 (top panel), GALAH DR3 (middle panel), and RAVE DR6 (bottom) survey datasets color-coded by the number density. From left to right, the panels are for \teff, \logg, [Fe/H] respectively. The numbers of the samples, medians and scatters of the differences of each stellar parameter are marked in each panel. And the residuals vary with each stellar parameter are shown below. The dashed lines are the reference lines.}
\label{figure12}
\end{figure*}

\begin{enumerate}

\item{APOGEE DR16}~\citep{2020ApJS..249....3A} published 473,307 median-high resolution (R $\sim$ 22,500) near-infrared spectra of 437,445 stars observed by APOGEE-north and APOGEE-south survey. Their stellar parameters and elemental abundances are derived by fitting the new version of MARCS templates~\citep{2020AJ....160..120J} and are calibrated based on photometry, astroseismology, and clusters. We cross-match our results with APOGEE DR16 and get a dataset of 11,842 LAMOST MRS spectra with APOGEE stellar labels with STARFLAG=0, ASPCAPFLAG=0, and PARAMFLAG=0 from the ASPCAP catalog, and Flag\_Teff=0, Flag\_logg=0, and Flag\_[Fe/H]=0 from LAMOST DR8 MRS \csn catalog.

\item{GALAH DR3}~\citep{2021MNRAS.506..150B} has published 768423 high resolution (R $\sim$ 28,000) optical spectra of 342,682 stars with stellar parameters (T$_{\rm{eff}}$, \logg, [Fe/H], $v_{mic}$, $v_{broad}$, $v_{rad}$) estimated by the model-driven Spectroscopy Made Easy (SME) and 1D MARCS model atmosphere. Also, astrometry from Gaia DR2 and photometry from 2MASS are employed to break the spectroscopic degeneracies and 30 different elements measured based on LTE/non-LTE computations. We cross-match our results with GALAH DR3 and get 21,015 LAMOST-\Rmnum{2} MRS spectra with corresponding stellar parameters after setting quality flags in both GALAH DR3 and LAMOST DR8 MRS \csn catalog.  

\item{RAVE DR6}~\citep{2020AJ....160...82S,2020AJ....160...83S} published 518,378 median-resolution (R $\sim$ 7500) spectra of 451,783 stars with two versions of stellar parameters derived by pipelines: MADERA (MAtisse and DEgas used in RAVE) and BDASP (Bayesian Distances Ages and Stellar Parameters). An asteroseismic-based calibration of stellar atmospheric parameters is used for giants. The abundance of the elements Fe, Al, and Ni, and an overall~\afe~ratio are determined with the pipeline GAUGUIN. We cross-match our results with RAVE DR6 and get a set of 833 LAMOST-\Rmnum{2} MRS spectra with corresponding RAVE DR6 stellar parameters after restricting the quality with the flags in RAVE DR6 and LAMOST DR8 MRS Cycle catalog. Here, we choose the latter version of the parameters for comparative analysis and do not use RAVE's chemical element abundance for analysis.

\end{enumerate}

The number density distributions of four surveys' datasets in the Kiel-diagram are shown in Fig.\ref{figure11}. For dwarfs, LAMOST MRS shows the same trend as GALAH and RAVE whose parameters are derived using Gaia parallax and photometric information, while APOGEE's dwarfs are different for that ASPCAP uses one asteroseismic calibration relation for warmer dwarfs and another approximate calibration based on isochrones for cooler dwarfs~\citep{2020AJ....160..120J}. For giant branches and red clumps (RC), four surveys show similar distributions.

The comparison of three stellar basic parameters between LAMOST MRS and the other three survey datasets is shown in Fig.~\ref{figure12}. In terms of effective temperature comparison, our results are most consistent with those of APOGEE, followed by GALAH, and finally RAVE. The scatters of the differences are contributed by both the error of LAMOST MRS and the comparison data sets. For \logg, the differences between LAMOST MRS gravity and GALAH, RAVE have small and similar dispersion ($\sim$ 0.1 dex), and the difference with APOGEE is larger, mainly because the former three all adopt the calibrations based on Gaia's parallax. The differences in metallicity between LAMOST MRS and APOGEE, GALAH are at the same levels, less than RAVE, which shows the advantage of high-resolution spectroscopy for calculating metal abundances.

\begin{figure*}[h]
\centering
\includegraphics[width=0.97\textwidth]{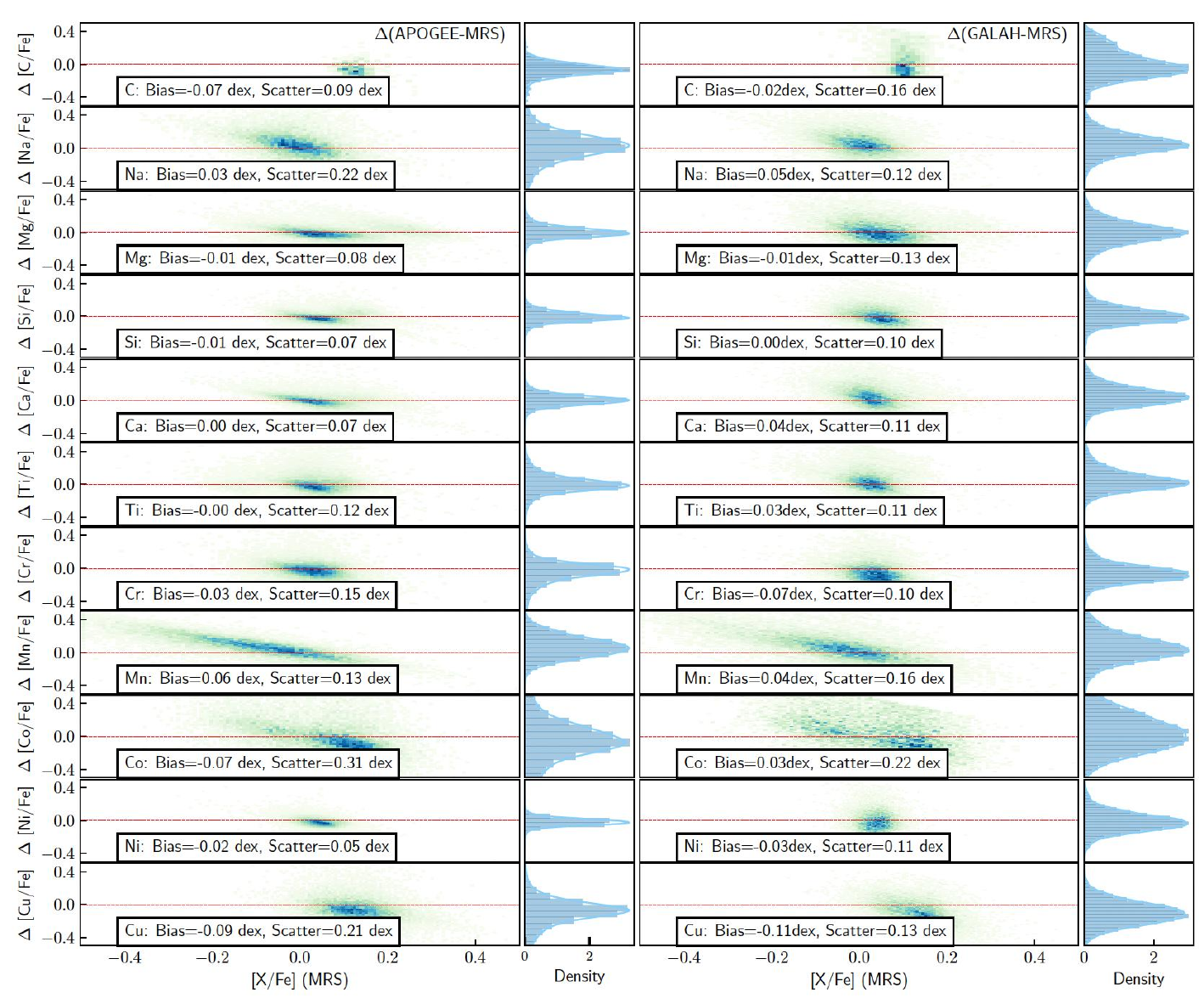}
\caption{Residuals density of elemental abundances between LAMOST DR8 MRS by \csn and APOGEE DR16 ASPCAP in left panels and GALAH DR3 in right panels. The distribution of the residuals is displayed beside. Biases and scatters of each element are marked in the text. The dashed line is the value of zero in each panel.}
\label{figure13}
\end{figure*}

Fig.\ref{figure13} shows the differences of 11 elements between LAMOST MRS and APOGEE DR16, GALAH DR3 very with [X/Fe]. Overall, the agreements with APOGEE DR16 seem better than GALAH DR3. The differences of $\alpha$ elements show less than those of [C/Fe], [Na/Fe], [Cr/Fe], [Co/Fe], and [Cu/Fe] which suffer large scatter and negative correlation trends with the elemental abundances. Because the blue bands of LAMOST MRS spectra are mainly dominated by Mg-\Rmnum{2} triplet and the red bands cover most Si lines. For the iron-peak element poor stars, most weak spectral characteristics in MRS spectra suffering median resolving power and low S/N make the measurements overestimated.

\subsection{Precision Analysis}

Precision refers to the degree of reproducibility of the measurement results. Among the sub-projects carried out by LAMOST MRS, the time-domain survey project is one of the very distinctive projects, many sources of which are repeatedly observed under different observed conditions. This provides a good opportunity for us to evaluate the precision of our results. Besides, the molecular cloud origin of the cluster led to the homogeneous chemical abundance pattern~\citep{2006AJ....131..455D,2011A&A...535A..30C,2016ApJ...817...49B}, which provides us a strategy to evaluate the precision of our metallicity and elemental abundances. Here, we analyzed repeated observations under different observing conditions in sub-section.\ref{subsub4-4} and studied elemental abundances in star clusters in sub-section.\ref{subsub4-5} to assess the precision of our measurements.

\begin{figure*}[htbp]
\centering
\includegraphics[width=0.95\textwidth]{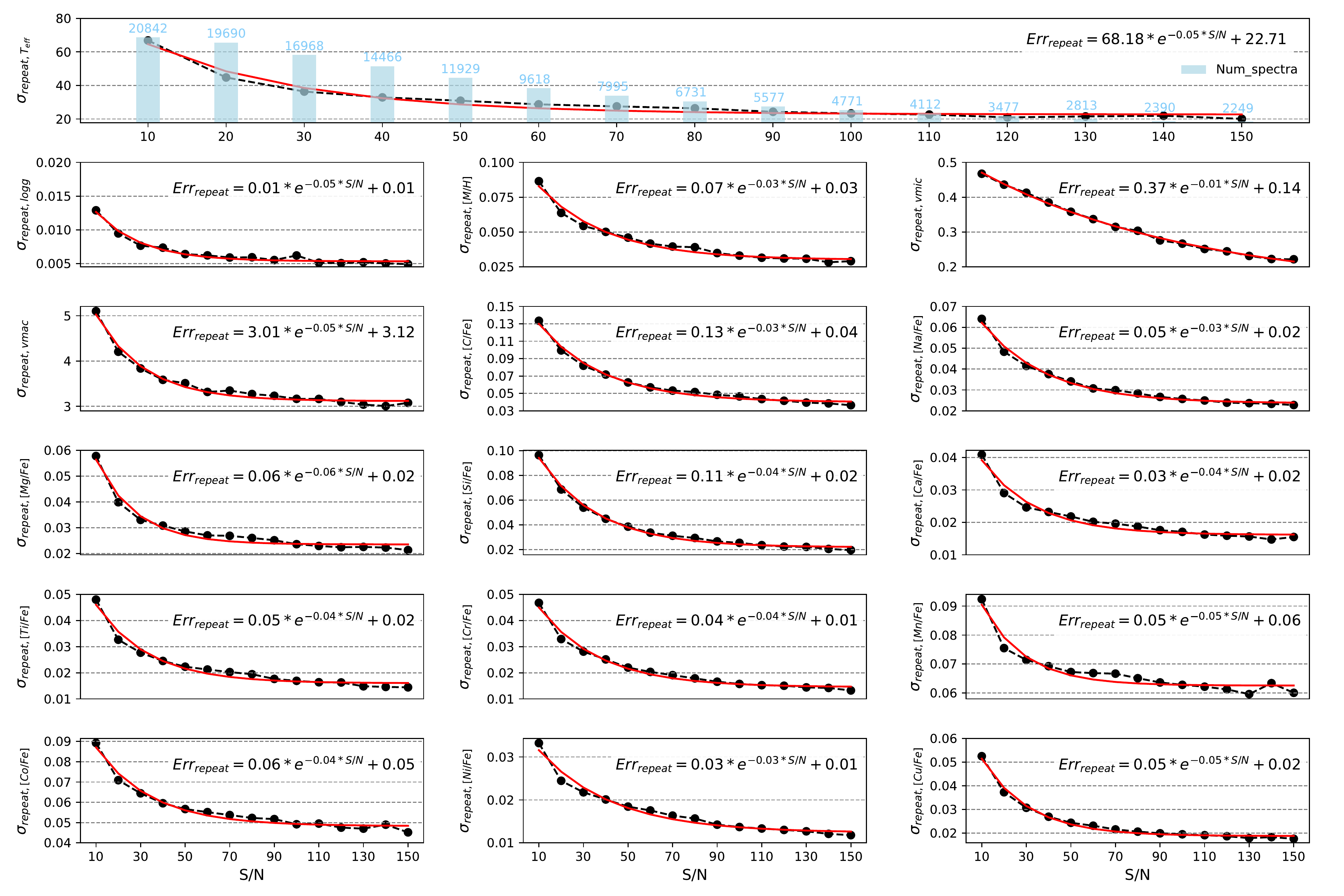}
\caption{Repeated errors of LAMOST MRS stellar parameters and chemical abundances derived by \csnn, plotted as a function of S/N. The 1-$\sigma$ values of the errors in each S/N bin are represented by black dots, and the fitting relationship is shown by a red curve with corresponding expression text in the upper right corner of each panel. In the top panel, we show the number of spectra in each S/N bin in light blue hist bars. Except for the [C/Fe], the number (ratio) of spectra in each S/N bin for other elements is basically the same as that shown in the top panel.}
\label{figure14}
\end{figure*}

\subsubsection{Precision assessed by repeated observations}
\label{subsub4-4}

We calculated the standard deviation of the repeated observation properties as precisions of stellar parameters and chemical element abundances based on LAMOST MRS sources repeat observed more than 6 times. Fig\ref{figure14} shows the trend of precision as a function of S/N and the number of spectra used in each S/N bin. It can be seen that the precision of the stellar labels decreases with the increase of the S/N, and tends to be stable after S/N higher than 100. For effective temperature, surface gravity, and metallicity, their precision is 76 K, 0.014 dex, and 0.096 dex at the S/N level of 10. When the S/N is increased to 100, the precision level converges at 22 K, 0.006 dex, and 0.043 dex. For the elemental abundances C, Na, Mg, Si, Ca, Ti, Cr, Mn, Co, Ni, and Cu their level of precision is 0.05 dex at S/N of 10 and eventually converges at 0.01 dex at S/N of 100. We fit the precision of each stellar parameter by an exponential relation: 
$$ 
Err_\textrm{repeat,x} = a*e^{-b*S/N} + c, 
$$ 
where $Err_\textrm{repeat,x}$ represents the precision of the stellar label $x$, the $S/N$ represents the signal-to-noise of the observation, and $a, b, c$ are the fitting coefficients (the values are in the Fig\ref{figure14}). 

For each observation, the error of the stellar label consists of two parts: one is the precision predicted by the relation of repeat observations with S/N; the other one is the fitting error contributed by the spectral fitting process. The final errors of the stellar parameters and chemical abundances are calculated following the equation: 
$$ 
Err_\textrm{final,x} = \sqrt{Err_\textrm{repeat,x}^2 + Err_\textrm{fit,x}^2}, 
$$ 
where $Err_\textrm{final,x}$ represents the final error of the stellar label $x$, $Err_\textrm{repeat,x}$ represents the repeated error and $Err_\textrm{fit,x}$ represents the fitting error.

\subsubsection{Precision validated by open clusters}
\label{subsub4-5}

\begin{figure*}[htbp]
\centering
\includegraphics[width=0.95\textwidth, angle=0]{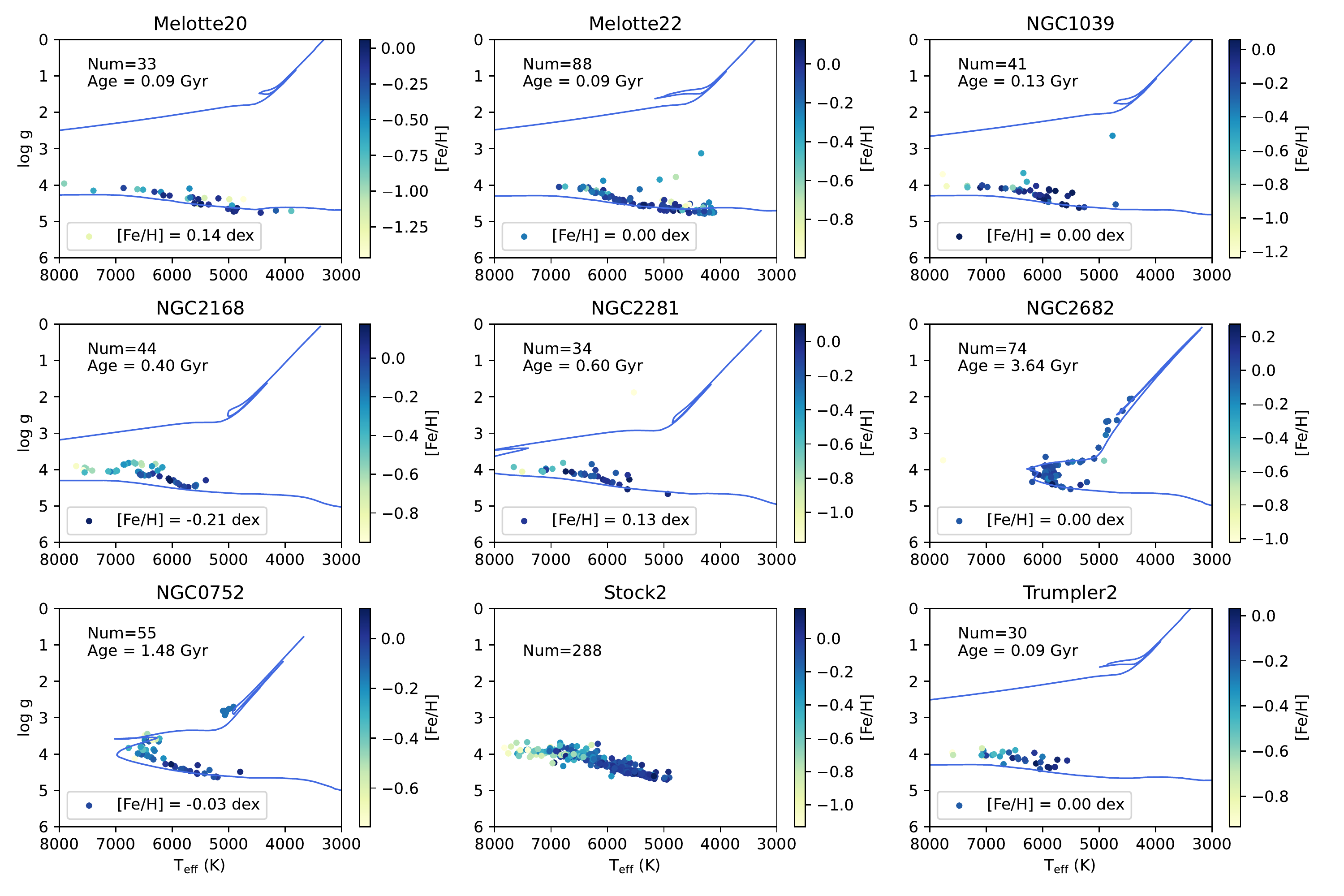}
\caption{Distribution of \csn estimates of nine open clusters in the Kiel-diagram, color-coded by their metallicities. The MIST isochrones are drawn based on the cluster's ages and metallicities provided by ~\cite{2019A&A...623A.108B, 2020MNRAS.499.1874M, 2021MNRAS.504..356D} except Stock 2.}
\label{Figure15}
\end{figure*}

\begin{figure*}[htbp]
\centering
\includegraphics[width=0.95\textwidth,height=0.3\textheight, angle=0]{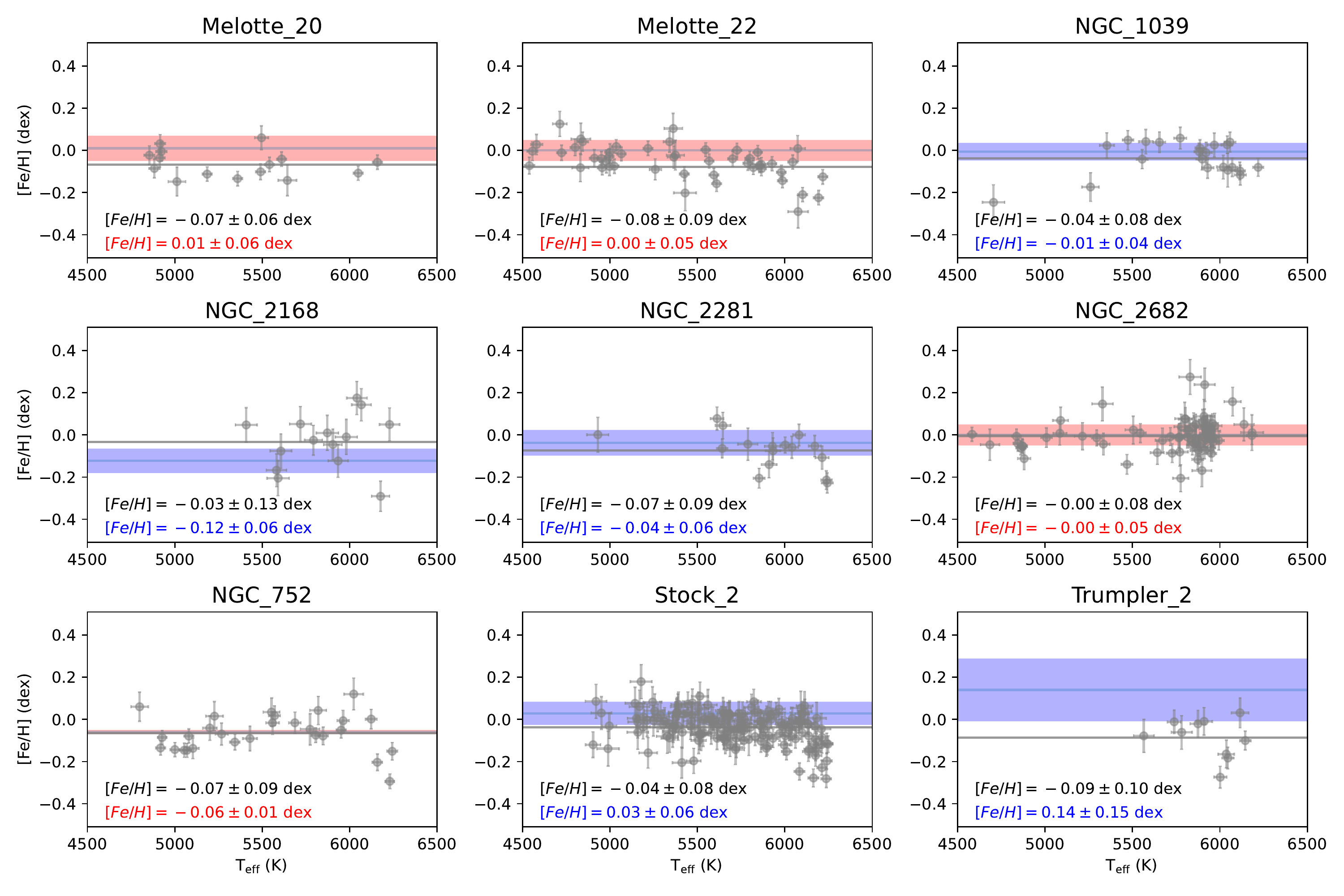}
\caption{Metallicities derived by \csn from LAMOST DR8 MRS spectra of 9 open clusters, shown as a function of their effective temperatures. If reported in the OCCAM survey catalog~\citep{2020AJ....159..199D}, the ranges of the clusters' metallicities are shown as red shadows, otherwise, blue shadows show the [Fe/H] ranges from the open clusters catalog~\citep{2019A&A...623A.108B,2020MNRAS.499.1874M,2021MNRAS.504..356D}. The means and dispersions are marked using the corresponding colors. The grey ones represent values calculated by LAMOST MRS samples in each panel.}
\label{Figure16}
\end{figure*}

\begin{figure*}[ht]
\centering
\includegraphics[width=0.9\textwidth,height=0.3\textheight, angle=0]{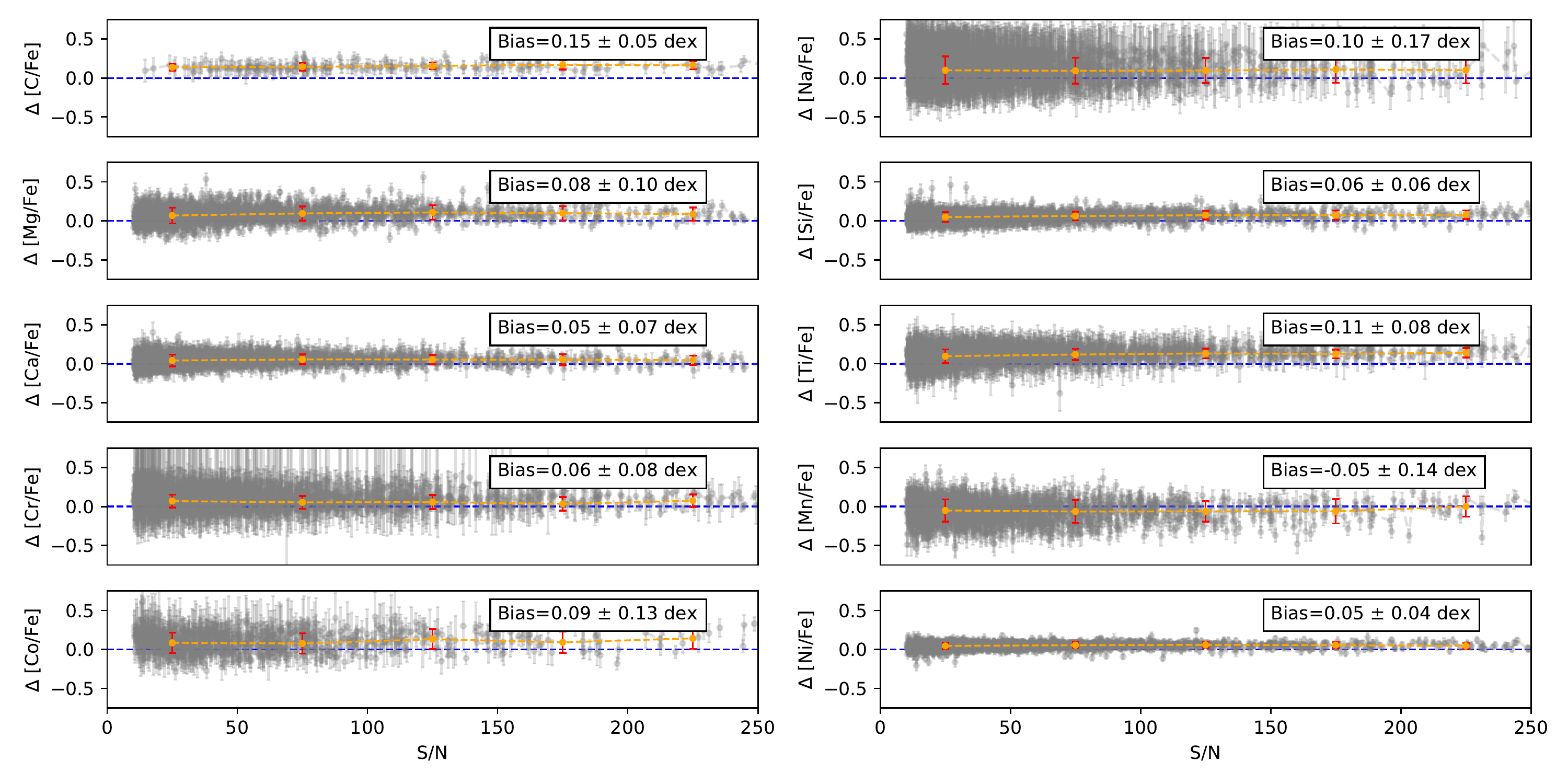}
\caption{Residuals between LAMOST MRS elemental abundances with respect to the reference values of the member stars of 9 clusters (ASCC 16, ASCC 19, ASCC 21, Melotte 20, Melotte 22, NGC 2632, NGC 2682, NGC 6811 and NGC 752), shown as a function of spectral S/N. The blue horizontal lines is the value of 0 in each panel. The error-bar of mean bias in S/N bins ([10, 50], [50, 100], [100, 150], [150, 200], [200, )) is shown in red. 1-$\sigma$ of the residuals represent its precision for each element in the legend.}
\label{Figure17}

\end{figure*}
\begin{figure*}[tb]
\centering
\includegraphics[width=0.9\textwidth,height=0.5\textheight,]{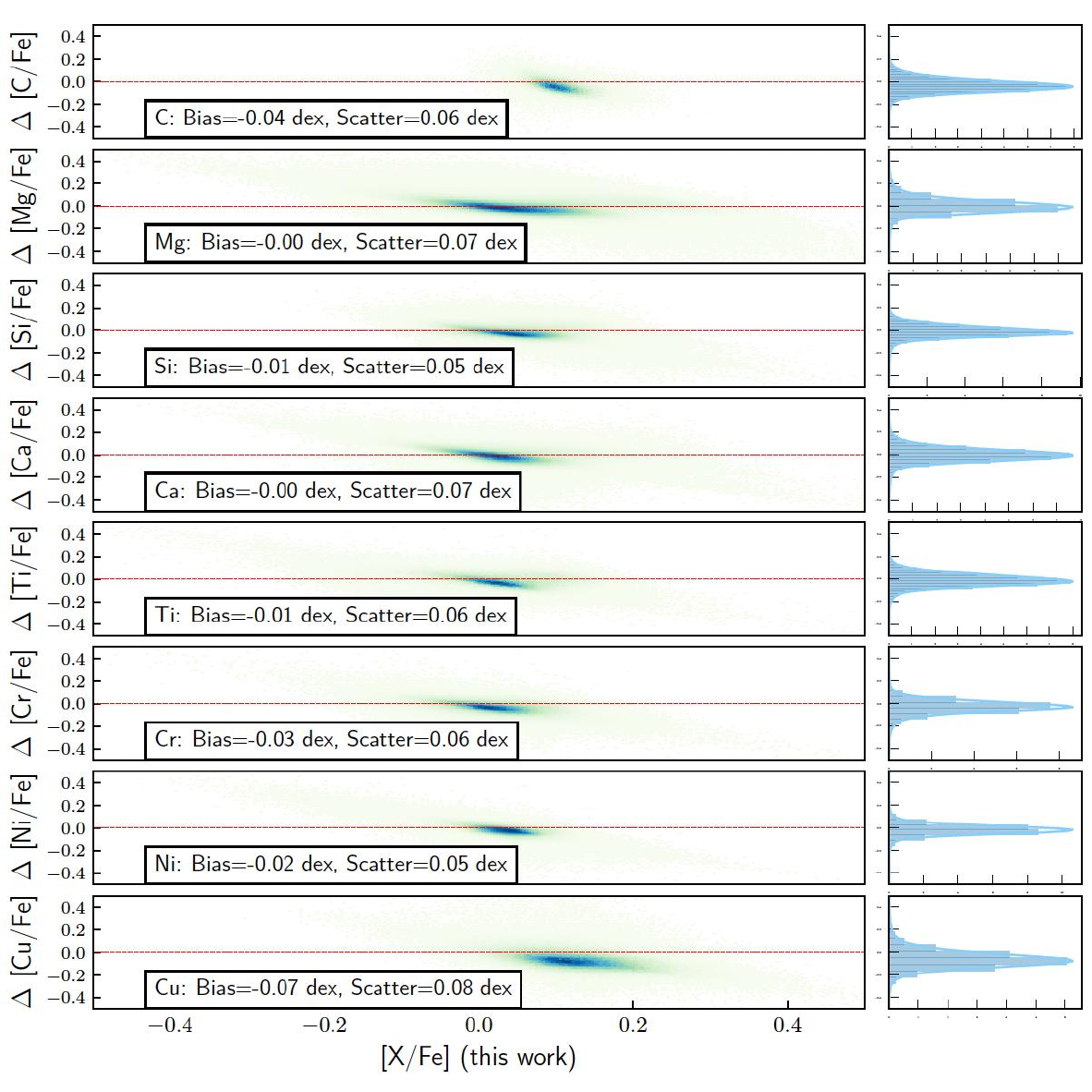}
\caption{Same as Fig.~\ref{figure13} but for the elements by \csnn~(this work) and SPCANet~\citep{2020ApJ...891...23W} as a function of elements [X/Fe]. The residual is represent SPECNet - \csnn~(this work).}
\label{figure18}
\end{figure*}

\citet{2018A&A...618A..93C} made use of Gaia DR2 astrometry and photometry information and obtained a catalog of the members and properties for 1229 clusters based on an unsupervised membership assignment algorithm.~\citet{2020A&A...633A..99C} expanded the number of the clusters to 1481.~\citet{2019A&A...623A.108B} derived parameters (age, distance modulus, and extinction) for 269 open clusters by applying a Bayesian tool to fit stellar isochrones to Gaia DR2 photometric data of the members.~\citet{2020MNRAS.499.1874M, 2021MNRAS.504..356D} presented a homogeneous catalog of fundamental parameters of 1743 open clusters in the Milky Way based on Gaia DR2 data. Cross-matching with the catalog of these open cluster members by a radius of 3 arcsecs, we obtain a raw sample of 3229 stars. To improve the reliability, we applied two conditions to the sample: 
$
4000 K < \textit{T}_\textrm{eff} < 7500~K~
\textrm{and}~ 
\textrm{Prob}_\textrm{member} > 0.75 
$ 
(the open clusters members catalog~\citep{2020A&A...633A..99C} provides a value for each star representing its probability belonging to the identified cluster). 891 objects are left with available stellar parameters after removing the eye-checked optical binaries based on their photometric images (same process as above in the last subsection). For each individual star with multiple observations, we retained the one with the highest $\textrm{S/N}_\textrm{blue}$ value and removed the rest repeated observations. 9 clusters with more than 30 members with available measurements of LAMOST MRS are finally selected, including Melotte 20, Melotte 22, NGC 1039, NGC 2168, NGC 2281, NGC 2682, NGC 752, Stock 2, and Trumpler 2. Totally, we got 687 samples of open cluster member stars used to derive the precision of the metallicity.

The effective temperatures and gravities derived by the LAMOST MRS spectra for the clusters' members are shown in Fig.\ref{Figure15}, in which the ages and metallicities of MIST isochrones are employed from~\citet{2019A&A...623A.108B,2020MNRAS.499.1874M,2021MNRAS.504..356D}. Our results (\teff, \logg, and [Fe/H]) fit the isochrones with the properties of the literature well except for individual stars, which can roughly indicate that most of the samples we selected are members of the target clusters. Their [Fe/H] as a function of \teff~are shown in Fig.\ref{Figure16}. For Melotte 20, Metlotte 22, NGC 2682, and NGC 752, we adopted the metallicities and dispersions by OCCAM survey~\citep{2020AJ....159..199D}, and for the rest five clusters, the values and dispersions of [Fe/H] are provided in the literatures~\citep{2019A&A...623A.108B,2020MNRAS.499.1874M,2021MNRAS.504..356D} as references. We also calculated the mean metallicity and dispersion of each cluster based on our results of the member stars. The dispersions are found to be consistent with the measurements provided by the OCCAM survey~\citep{2020AJ....159..199D} and the literature, although biases exist for the zero points to some clusters, especially for the Trumpler 2 suffering 0.23 dex bias. For most clusters, their [Fe/H] dispersions are below 0.1 dex.

For each chemical abundance, we used the cluster member stars with LAMOST elements comparing with the reference values of the clusters by OCCAM survey~\citep{2020AJ....159..199D}, and took the standard deviations of the weighted biases (calculation process same as the accuracy of stellar parameters above) as the precision.

The OCCAM survey~\citep{2020AJ....159..199D,2022AJ....164...85M}, based on the data collected by APOGEE, specifically focuses on analyzing the ages, distances, reddening, and chemical abundances of stars in open clusters to gain insights into the formation and evolution of the Milky Way and its stellar populations. The OCCAM DR17 catalog contains 153 clusters properties based on 26,699 APOGEE observations. We got 3720 LAMOST MRS spectra (S/N $\geq$ 10) of 52 clusters from which we selected 9 open clusters (ASCC 16, ASCC 19, ASCC 21, Melotte 20, Melotte 22, NGC 2632, NGC 2682, NGC 6811 and NGC 752) by filtering the number of the cluster members observed by LAMOST larger than 50.

The residuals between LAMOST MRS elemental abundances with respect to the reference values of the members of 9 open clusters are displayed in Fig.\ref{Figure17}. It should be noted that the OCCAM DR17 catalog does not provide copper abundance information, so no copper precision can be provided here. Although plotted error bars of the residuals in S/N bins in the figure, we did not provide the precision by clusters at each S/N level because the number of samples averaged in each S/N bin is insufficient to provide statistically significant values. Additionally, the precision does not vary significantly as the S/N changes. Finally, we got the overall precision is 0.05 dex, 0.17 dex, 0.10 dex, 0.06 dex, 0.07 dex, 0.08 dex, 0.08 dex, 0.14 dex, 0.13 dex, and 0.04 dex of elements (C, Na, Mg, Si, Ca, Ti, Cr, Mn, Co and Ni). The results of LAMOST MRS show different homogeneity within acceptable dispersion for different clusters, a large zero-point shift of C, and a large scatter for Na. The homogeneity illustrates LAMOST MRS data can contribute to the study of elements in star clusters. We displayed the elemental abundances of several clusters calculated using LAMOST MRS observations in the APPENDIX.

\subsection{The Contents of the LAMOST-\Rmnum{2} MRS Value-added Catalogue}

This work provides a catalog of stellar parameters and chemical elements (C, Na, Mg, Si, Ca, Ti, Cr, Mn, Co, Ni, and Cu) of 1.38 million LAMOST-\Rmnum{2} MRS spectra. Although not the first publication of chemical abundances for LAMOST MRS, it is the first one based on the ab initio atmospheric model and the first application of Cycle-StarNet to more than 1 million real data sets. The method incorporates the physical model and data-driven approaches to improve the shortcomings of the low signal-to-noise ratio of observed spectra and imperfections of theoretical synthetic spectra in the fitting process. The catalog is available online and contains information: the identifier for the corresponding star (starid), LAMOST spectrum identifier ($\rm{medid}$), Gaia identifier for 69 percent of spectra ($Gaia\_source\_id$), coordinate information (right ascension (RA), declination (Dec)), signal-to-noise of the spectra (S/N of blue and red arm), effective temperature (\teff), surface gravity (\logg), metallicity [Fe/H], macroturbulence, microturbulence, elemental abundance ([X/Fe]), their errors and quality flags, and the total quality flag reported. A description of columns of the catalog is shown in Tab~\ref{Tab:description} and the full catalog can be accessed online at \url{http://paperdata.china-vo.org/LAMOST/LAMOST_DR8_MRS_CSN_parameters_abundances.csv}.

\begin{table*}[htp]
\caption{Description of the columns of LAMOST DR8 MRS stellar parameters and chemical abundances catalog.}
\begin{center}{}
\label{Tab:description}
\begin{tabular}{lll}
\hline\noalign{\smallskip}
\hline\noalign{\smallskip}
Col. &  Name  &  Description \\
\hline\noalign{\smallskip}
1 &$\rm{starid}$ & ID for corresponding star based on the R.A. and decl., with the form of ``LAMOST Jddmmss ddmmss"\\
2 &$\rm{medid}$ & LAMOST spectral ID, inform of Date-PlateID-SpectrographID-FiberID-MJM-PiplineVersion\\
3 &Gaia$\_{\textrm{source}\_\textrm{id}}$&Gaia source id by cross-matching Gaia DR2\\
4 &$\textrm{RA}$ &  Right ascension of J2000 ($^{\circ}$) \\
5 &$\textrm{Dec}$ & Declination of J2000 ($^{\circ}$) \\
6 &$\textrm{S/N}\_\textrm{blue}$ & Signal-to-noise of the blue arm \\
7 &$\textrm{S/N}\_\textrm{red}$  & Signal-to-noise of the red arm \\
8 &$\textit{T}_{\textrm{eff}}$ & Effective temperature (K) \\
9 &\logg& Surface gravity (dex) \\
10 &[Fe/H] & Metallicity with respect to hydrogen (dex) \\
11 &$v_{mac}$ & Macroturbulence (km/s) \\
12 &$v_{mic}$ & Microturbulence (km/s) \\
13-23 &[X/Fe]  & elemental abundance with respect to iron (dex) \\
24 &Err$\_\textit{T}_{\textrm{eff}} $ & Error of the $\textit{T}_{\textrm{eff}}$ (K) \\
25 &Err\_logg & Error of the \logg~(dex) \\
26 &Err\_[Fe/H] & Error of the metallicity (dex) \\
27 &Err\_vmac & Error of the macroturbulence (km/s) \\
28 &Err\_vmic & Error of the microturbulence (km/s) \\
29-39 &Err\_[X/Fe]  & Error of the elemental abundance with respect to iron (dex) \\
40 &Flag$\_\textit{T}_{\textrm{eff}}$ & Quality flag of the $\textit{T}_{\textrm{eff}}$: 0 for good, while 1 for bad \\
41 &Flag\_logg & Quality flag of the \logg: 0 for good, while 1 for bad \\
42 &Flag\_[Fe/H] & Quality flag of the metallicity: 0 for good, while 1 for bad \\
43 &Flag\_vmac & Quality flag of the macroturbulence: 0 for good, while 1 for bad \\
44 &Flag\_vmic & Quality flag of the microturbulence: 0 for good, while 1 for bad \\
45-55 &Flag\_[X/Fe] & Quality flag of the elemental abundance with respect to iron: 0 for good, and 1 for bad \\
56 &Flag\_Quality & Quality flag of the results: 0 for good, 1 for bad spectra, 2 for large fitting k-square value ($\le$ 1.5), \\
 & & 3 for without \logg~calibrated, and 4 for suspected binary or multi-system. \\
\noalign{\smallskip}\hline
\end{tabular}
\end{center}
\tablecomments{The full catalog can be accessed on-line.}
\end{table*}

Besides, a comparison of eight elements (C, Mg, Si, Ca, Ti, Cr, Ni, and Cu in common) with the updated version of the data-driven catalog provided by \citet{2020ApJ...891...23W} are displayed in Fig.~\ref{figure18}. For most stars, the residuals of the elements (Mg, Si, Ti, Cr, Ni) obtained by two methods are around 0 with dispersion less than 0.06 dex, while the residuals of C and Ca have different slopes of trends. For [Cu/Fe], a bias of -0.08 dex exists with a slope when [Cu/Fe] is higher than 0.2 dex, which also appears in comparison with APOGEE.

\section{Summary}

The traditional measurement of chemical abundances is mainly for high-resolution spectroscopy. For individual elements, selecting the spectral lines of the special window to fit the theoretical model or calculating the equivalent width (EW) of the characteristic lines to derive the element abundances. It often requires relatively clean absorption line features, which further require a high resolving power and a high signal-to-noise ratio of a spectrum. For medium-resolution spectra, the characteristic lines of individual elements always suffer contamination and blend. The full spectrum fitting provides the possibility to extract chemical information from the medium-/low-resolution spectra, although the precision may not be so high and the accuracy needs to be further calibrated by benchmark stars or clusters.

We derived the fundamental stellar parameters (\teff, \logg, [Fe/H], $v_{mic}$ and $v_{mac}$) and 11 chemical abundances ([C/Fe], [Na/Fe], [Mg/Fe], [Si/Fe], [Ca/Fe], [Ti/Fe], [Cr/Fe], [Mn/Fe], [Co/Fe], [Ni/Fe], and [Cu/Fe]) for 1.38 millions FGKM-type stars of the Medium-Resolution Spectroscopic Survey (MRS) from LAMOST-\Rmnum{2} DR8 by fitting MARCS theoretical synthetic spectra. We use a domain adaptation method to reduce the gap between the observed and synthetic spectra and realize the measurement of 5 stellar parameters and 11 abundances simultaneously. The surface gravities of the stars are calibrated after the bolometric luminosities are estimated based on the 2MASS photometric data and Gaia parallaxes.

The accuracy of the stellar labels is evaluated by the PASTEL catalog, asteroseismic \logg~dataset, and three other spectroscopic surveys (APOGEE DR16, GALAH DR3, and RAVE DR6). The accuracy can reach 150 K to the PASTEL catalog for \teff, 0.11 dex to the asteroseismic samples for \logg, and 0.15 dex to the PASTEL catalog for [Fe/H]. Our results show consistency to APOGEE, GALAH, and RAVE with minor biases and different degrees of scatters.

The precision is assessed using repeated observations and validated by open cluster members. For \teff, \logg~and [Fe/H], their precision can achieve 76 K, 0.014 dex, and 0.096 dex at the S/N level of 10. When the S/N is increased to 100, the precision level converges at 22 K, 0.006 dex, and 0.043 dex. For the elemental abundances C, Na, Mg, Si, Ca, Ti, Cr, Mn, Co, Ni, and Cu their precision is 0.06-0.2 dex at S/N of 10 and eventually converges at 0.01 dex at S/N of 100. The precision of LAMOST MRS open cluster members are show scatters of about 0.04-0.10 dex for elements except for Na, Mn, and Co.

A full catalog is provided online at \url{http://paperdata.china-vo.org/LAMOST/LAMOST_DR8_MRS_CSN_parameters_abundances.csv}. In the future, the age information of LAMOST MRS stars can be derived based on these stellar atmospheric parameters. Combined with kinematic parameters, we will analyze the chemical kinematic evolution of the Milky Way in the next work.

\software{
Numpy~\citep{oliphant_guide_2006},
Scipy~\citep{jones2001scipy},
Matplotlib~\citep{Hunter:2007},
Pandas~\citep{pythonpandas},
Astropy~\citep{astropy:2018},
PyTorch~\citep{paszke2017automatic},
iSpec~\citep{2014A&A...569A.111B,2019MNRAS.486.2075B},
Apache Spark~\citep{Zaharia2016},
Topcat~\citep{2005ASPC..347...29T}
}

\acknowledgments

We thank the cluster catalog supplied by Dr. W. S. Dias and advices from Dr. Sven Buder, Dr. Chao Liu, and Dr. Xiao-Ting Fu.

This work was funded by the National Natural Science Foundation of China (Grant No.12103068, 11390371, 11703053, 11973060, 12090044), the National Key R\&D Program of China (Grant No.2019YFA0405102), the Cultivation Project for LAMOST Scientific Payoff and Research Achievement of CAMS-CAS and the Science Research Grants from the China Manned Space Project with NO.CMS-CSST-2021-B05. Y.S.T. acknowledges financial support from the Australian Research Council through DECRA Fellowship DE220101520.

Guoshoujing Telescope (the Large Sky Area Multi-Object Fiber Spectroscopic Telescope LAMOST) is a National Major Scientific Project built by the Chinese Academy of Sciences. Funding for the project has been provided by the National Development and Reform Commission. LAMOST is operated and managed by the National Astronomical Observatories, Chinese Academy of Sciences.

\bibliography{reference}
\appendix

\begin{figure}[b]
  \centering
  \includegraphics[
  width=0.95\textwidth, 
  angle=90]{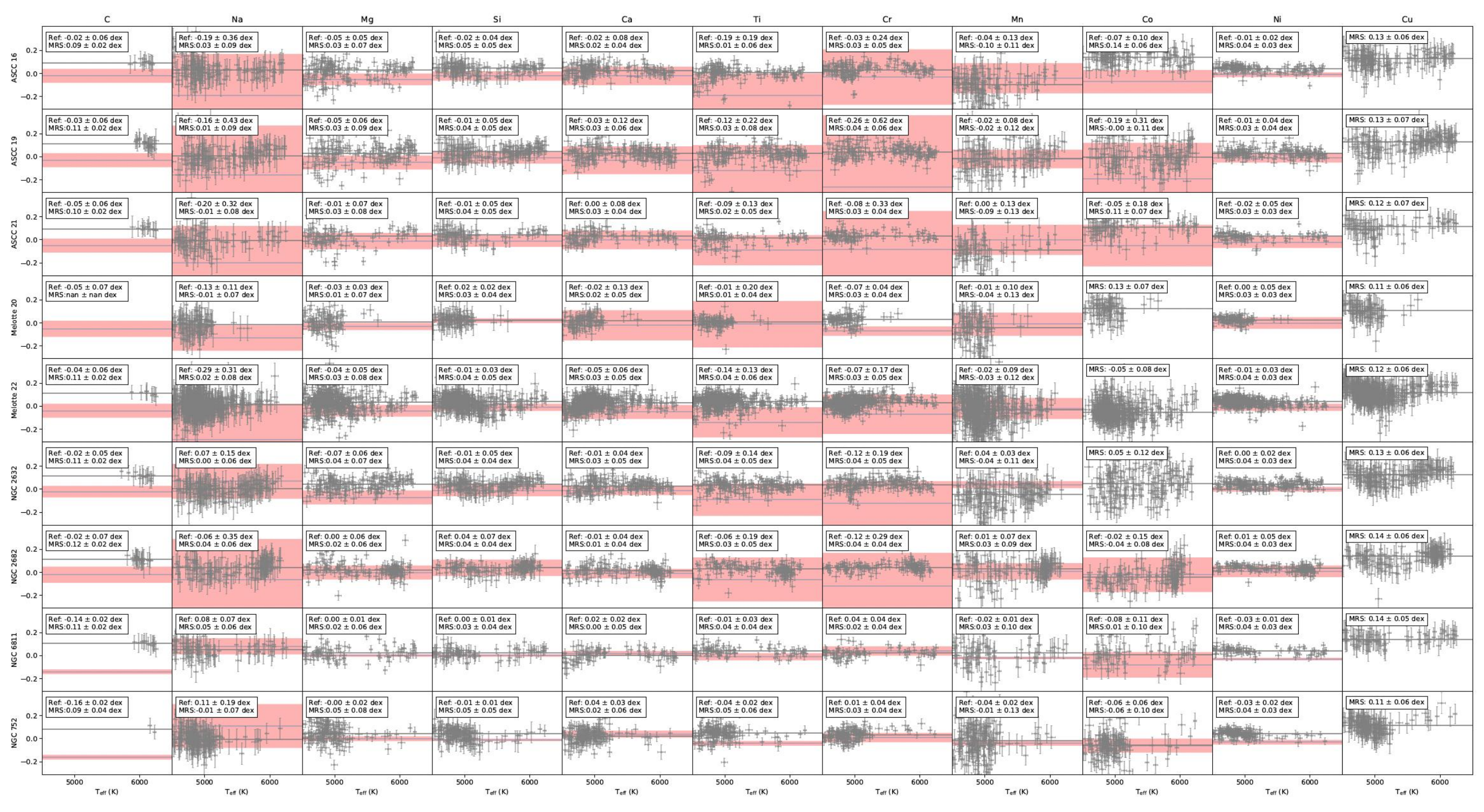}
  \caption{The elemental abundances by \csnn displayed by grey points with errors as a function of \teff~of the clusters (Melotte 20, Melotte 22, NGC 2682, NGC 752, NGC 1039, NGC 2168, NGC 2281 and Stock 2), of which has reference abundances range from the OCCAM DR17 catalog shown in red shadow.}
  \label{fig:Figure-a}
\end{figure}

Stellar clusters provide an excellent means to evaluate the precision of chemical abundance measurements. We utilized nine stellar clusters to evaluate the precision of chemical abundance measurements in LAMOST MRS spectra (in Figure.~\ref{fig:Figure-a}). These clusters serve as ideal samples for assessment, as their member stars share similar ages, compositions, and distances. By comparing the variations in chemical abundances among stars within these clusters to the measurement uncertainties, we can gain insights into the precision and reliability of LAMOST MRS measurements.

\end{document}